\documentclass[8pt]{article}
\usepackage[a4paper, total={16cm, 22cm}]{geometry}
\usepackage[utf8]{inputenc}
\usepackage{graphicx,wrapfig}
\usepackage{float}
\usepackage{amsmath}
\usepackage{amssymb}
\usepackage{comment} 
\usepackage{color}
\usepackage{bm}
\usepackage{titlesec}
\usepackage{titling}
\usepackage{authblk}

\definecolor{cadetgrey}{rgb}{0.57, 0.64, 0.69}

\usepackage{hyperref}
\hypersetup{
	colorlinks=true,       
	linkcolor=black,          
	citecolor=cadetgrey,        
}

\newcommand*\diff{\mathop{}\!\mathrm{d}}
\usepackage[utf8]{inputenc}

\newcommand*\samethanks[1][\value{footnote}]{\footnotemark[#1]}

\title{{\textbf {Quality of internal representation shapes learning performance in feedback neural networks}}}
\date{}

\author[1,2]{Lee Susman \thanks{Equal contribution, listed in random order}}
\author[3]{Francesca Mastrogiuseppe  \samethanks}
\author[2,4]{Naama Brenner \thanks{Equal contribution}}
\author[2,5]{Omri Barak \samethanks}
\affil[1]{Interdisciplinary program in Applied Mathematics, Technion Israel Institute of Technology, Haifa, 32000, Israel}
\affil[2]{{Network Biology Research Laboratories, Technion Israel Institute of Technology, Haifa, 32000, Israel}, \newline {\normalfont{lee.susman@gmail.com}}}
\affil[3]{Gatsby Computational Neuroscience Unit, University College London, London \newline {\normalfont{fran.mastrogiuseppe@gmail.com}}}
\affil[4]{Dept. of Chemical Engineering, Technion Israel Institute of Technology, Haifa, 32000, Israel, \newline {\normalfont{nbrenner@technion.ac.il}}}
\affil[5]{Rappaport Faculty of Medicine, Technion Israel Institute of Technology, Haifa, 32000, Israel, \newline {\normalfont{omri.barak@gmail.com}}}

\begin{document}
	
	\maketitle
	
	
	\vskip -2cm

	\begin{abstract}
		
		A fundamental feature of complex biological systems is the ability to form feedback interactions with their environment.
		A prominent model for studying such interactions is reservoir computing, where learning acts on low-dimensional bottlenecks.
		Despite the simplicity of this learning scheme, the factors contributing to or hindering the success of training in reservoir networks are in general not well understood.
		In this work, we study non-linear feedback networks trained to generate a sinusoidal signal, and analyze how learning performance is shaped by the interplay between internal network dynamics and target properties. 
		By performing exact mathematical analysis of linearized networks, we predict that learning performance is maximized when the target is characterized by an optimal, intermediate frequency which monotonically decreases with the strength of the internal reservoir connectivity. 
		At the optimal frequency, the reservoir representation of the target signal is high-dimensional, de-synchronized, and thus maximally robust to noise. 
		We show that our predictions successfully capture the qualitative behaviour of performance in non-linear networks. 
		Moreover, we find that the  relationship between internal representations and performance can be further exploited in trained non-linear networks to explain  behaviours which do not have a linear counterpart.
		Our results indicate that a major determinant of learning success is the quality of the internal representation of the target, which in turn is shaped by an interplay between parameters controlling the internal network and those defining the task.
		
	\end{abstract}
	
	\section{Introduction}
	
	A fundamental feature of the brain, and biological networks in general, is the ability to form closed-loop interactions with their environment. 
	Such interactions are often implemented through a dimensionality bottleneck: while networks typically consist of large numbers of units, signals exchanged with the environment are low-dimensional. 
	In fact, external stimuli can often be represented in terms of a few scalar variables (e.g.~the angle and speed of a tennis ball approaching); these low-dimensional variables are encoded in the high-dimensional activity of a large population of neurons \cite{chaudhuri2019intrinsic, Low2018} before being again transformed into low-dimensional decision variables and motor outputs (e.g.~the angle and speed of the hand holding the racket).
	
	Simple but effective models for studying closed-loop interactions are feedback networks. 
	These models implement a simple form of closed-loop interaction: the output (or readout) signal, which is extracted from a reservoir of randomly connected units as a linear combination of unit activities, is directly injected back into the reservoir as external input \cite{JaegerHaas, Maass2007}.
	By adjusting the weights which specify how reservoir activity is mapped to the output, feedback networks can be trained to produce the desired readout signal. 
	In the most common training algorithms \cite{Jaeger, LukoseviciusJaeger, SussilloAbbott}, readout weights are updated through least-squares (LS) regression; this can be performed only once, by using a complete batch of activity samples \cite{Jaeger}, or in an online fashion, by recursively integrating activity samples as they are simulated \cite{SussilloAbbott, Jaeger2002}.
	
	What kind of closed-loop dynamics can feedback networks implement?
	Despite some theoretical advancement \cite{Maass2007, MassarMassar, RivkindBarak, MastrogiuseppeOstojic3,  Schuessler2020}, computational properties of feedback networks are still poorly understood.
	Early theoretical work has indicated that most feedback models are expected to be able to approximate readout signals characterized by arbitrarily complex dynamics \cite{Maass2007}. 
	However, it has been reported that not all feedback architectures and target dynamics result in the same performance: trained networks can experience dynamical instabilities \cite{RivkindBarak, MastrogiuseppeOstojic3}, and converge to fragile solutions for certain choices of the feedback architecture and parameters \cite{SussilloAbbott, Logiaco2019}. 
	
	For a fixed task, several studies have reported that training performance is strongly influenced by the overall strength of recurrent connections in the reservoir \cite{LegensteinMaass, SussilloAbbott,LukoseviciusJaeger}.
	Specifically, performance is high when recurrent connections are strong, but not strong enough to lead to the appearance of chaotic activity \cite{Sompolinsky1988}  -- a parameter region named \emph{edge-of-chaos} \cite{Bertschinger2004}. Intuitively, the edge-of-chaos defines an optimal tradeoff point where the internal reservoir dynamics are rich but stable.
	
	Reservoir activity, however, is not determined by connectivity alone: because the system is coupled to the environment, activity depends also on the statistics and dynamics of the target signal, which specify the task.
	How the internal reservoir dynamics interact with the target in determining trained networks performance is a fundamental question in feedback systems which is still not well understood \cite{JaegerRiddle}.
	In particular: are there specific target features which optimize performance, and how do they depend on internal properties of the reservoir network? For given values of the target parameters, what are the properties of reservoir activity that support optimal training? 
	How sensitive is the optimal performance to the learning algorithm?
	To the current date, these questions remain largely unsolved.
	
	In this work, we consider a simple setup consisting of a non-linear reservoir of rate units which is trained to sustain a sinusoidal output with given frequency $\omega$.
	Consistently across three different training techniques, we find that learning performance is maximized at a finite ``preferred'' frequency $\bar{\omega}$, which in turn depends on  reservoir connectivity: as the connectivity strength is increased towards the edge of chaos, $\bar{\omega}$ decreases towards zero. This nontrivial dependence of performance, even in a simple task, provides a test case to study the interplay between reservoir and target properties and its effects on learning.
	
	To gain analytical insight into this phenomenon, we consider a simplified setup where reservoir dynamics are linearized, and perform exact mathematical analysis.
	By averaging over the ensemble of random reservoir networks, we characterize reservoir activity in response to the target signal, and show that a ``resonance'' frequency $\omega^*$ emerges, which decreases with the connectivity strength.
	Under this frequency, dimensionality of neural activity is maximal and synchrony across different units in the reservoir is minimal. 
	When training the network to output the target signal, feedback interactions are most robust in the vicinity of the resonance frequency, thus resulting in optimal performance.
	Moreover, this behaviour is predicted to be qualitatively consistent across different training algorithms, even if  performance itself is sensitive to the algorithm used. 
	We show that our theoretical predictions correctly capture the qualitative behaviour of learning performance observed numerically in non-linear network models. 
	Overall, our results shed light on the learning capacity of recurrent network architectures by quantifying how learning precision is determined by the interaction between internal reservoir connectivity and target dynamics.

	
	\section{Results}
	
	
	\subsection{Emergence of a preferred frequency in trained feedback networks}
	\label{sec:nonlinear_phenomenon}
	We consider a reservoir network consisting of $N$ units characterized by the evolution dynamics:
	\begin{equation}\label{eq:dyn_reservoir}
	\dot{\mathbf{x}}(t) = -\mathbf{x}(t) + \mathbf{J}\Phi(\mathbf{x}(t)) + \mathbf{m} u(t)
	\end{equation}
	where $\Phi(x) = \tanh (x)$ is applied to the activation vector $\mathbf{x}$ element-wise. 
	Recurrent weights $\mathbf{J}$ are fixed, and are generated independently from the normalized Gaussian distribution $\mathcal{N}(0,g^2/N)$ \cite{Sompolinsky1988, SussilloAbbott}, so that the parameter $g$ controls the strength of reservoir connectivity. 
	The one-dimensional external signal $u(t)$ acts as a forcing on the reservoir through input weights $\mathbf{m}$, which are fixed and drawn as independent standard Gaussian variables.
	
\begin{figure}
		
		\centering
		\includegraphics{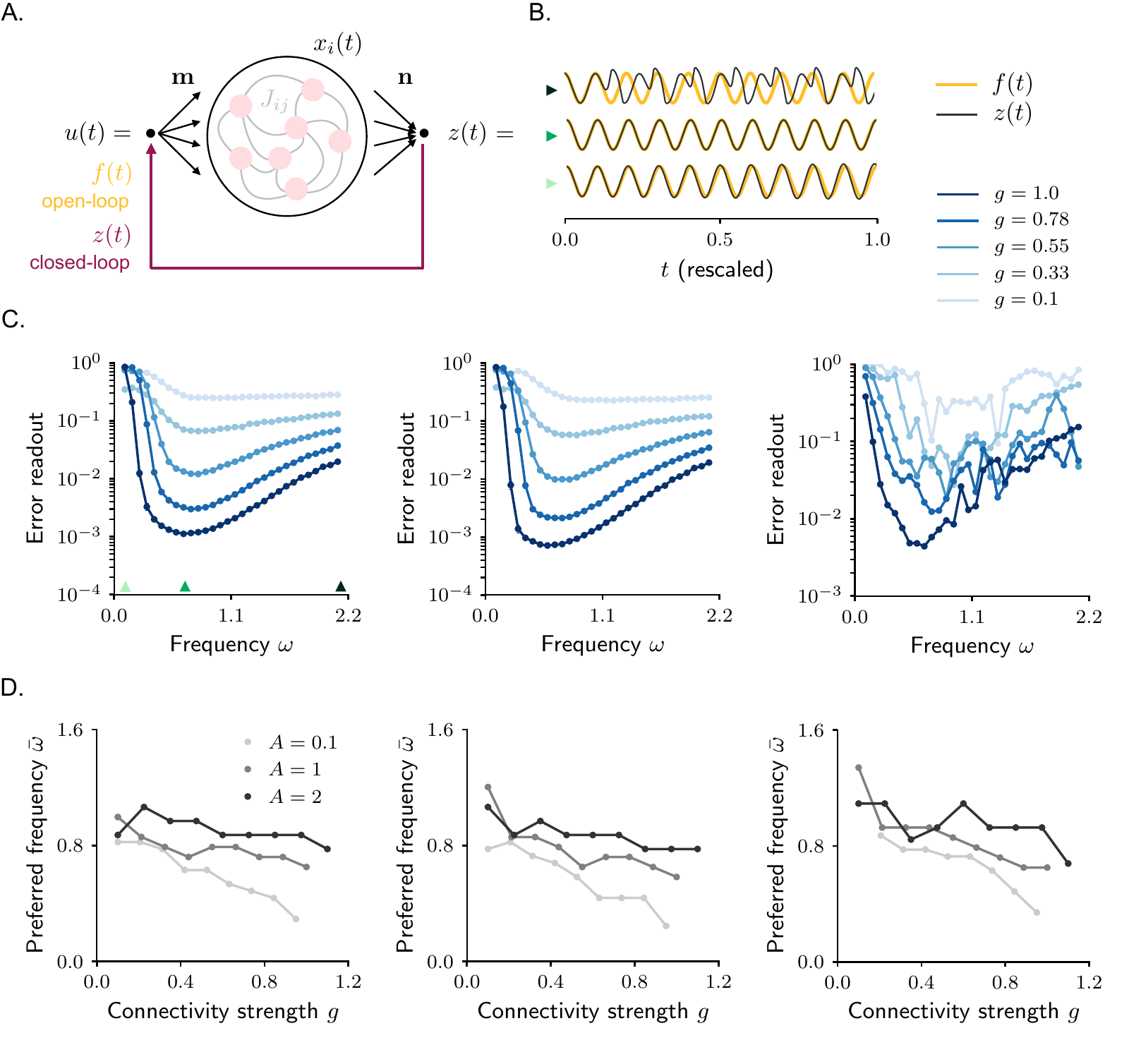}
		\caption{
			\textbf{Emergence of preferred frequency in non-linear feedback networks trained to sustain a sinusoidal output.}
			{\bf A.} Illustration of network architecture used in open (yellow) and closed (purple) loop.
			{\bf B.} Example readout signal (dark grey; target shown in yellow) for three learning trials corresponding to the frequency values indicated in green in {\bf C} ($w=$ 0.1, 0.7, 2.1; $g=1$). 
			Other parameters as in {\bf C}, except (for illustration purposes) training is performed on a smaller number of target cycles ($N^{\text{tot}}=4$ and $N^{\text{tr}}=2$, see Appendix \ref{app:training}). LS regression was used for training; examples trials for Ridge and RLS are reported in Supp.~Fig.~\ref{fig:S2}.
			{\bf C.} Readout error as a function of $\omega$, for a range of $g$ values (blue shades), for networks trained via LS (left), ridge regression (middle) and RLS (right). We take here $A=1$. Training details and parameters are reported in Appendix \ref{app:training}.
			{\bf D.} Error-minimizing frequency $\bar{\omega}$ as a function of $g$, for the three learning algorithms as in {\bf C}. Three different target amplitudes $A$ were tested (grey shades).
			\label{fig:fig1}
		}
	\end{figure}

	The output of the reservoir network is a one-dimensional readout signal, defined as:
	\begin{equation}
	z(t) = \mathbf{n}^\top \Phi(\mathbf{x}(t))
	\end{equation}
	through a set of decoding weights $\mathbf{n}$ that are assumed to be plastic. 
	The feedback is realized by using the output signal as input: $u(t) = z(t)$  (see Fig.~\ref{fig:fig1}{\bf A} for an illustration), which yields the final autonomous  dynamics
	\begin{equation}\label{eq:dyn_feedback}
	\dot{\mathbf{x}}(t) = -\mathbf{x}(t) + (\mathbf{J} + \mathbf{m} \mathbf{n}^\top) \Phi(\mathbf{x}(t)).
	\end{equation}
	During training, the vector $\mathbf{n}$ is updated until the output $z(t)$ best matches the desired target $f(t)$. 
	The target function that we consider is a simple sinusoidal wave of frequency $\omega$, i.e. $f(t) = A \cos(\omega t)$.
	
	We trained multiple instances of this feedback architecture and analyzed how performance depends on the frequency of the target signal $\omega$ and on internal coupling strength $g$ (Fig.~\ref{fig:fig1}).
	Three common training algorithms (least-squares (LS) regression, ridge regression \cite{HoerlRidge} and recursive least-squares (RLS) \cite{Liu, SussilloAbbott}) were used (training details are reported in Appendix \ref{app:training}). 
	We quantified the error as the mismatch between the target $f(t)$ and the readout $z(t)$ averaged over a finite number of target cycles in the post-training activity.
	
	We observe that, for fixed reservoir connectivity $g$, the accuracy of signal reconstruction by the output strongly depends on the target frequency: 
	while training on one frequency results in highly precise readout for many cycles, others result in a runaway from the target signal (Fig.~\ref{fig:fig1}{\bf B}).
	For every value of $g$, the error has a non-monotonous dependence on $\omega$, and reaches a minimum at a finite frequency that we name $\bar{\omega}$ (Fig.~\ref{fig:fig1}{\bf C}). 
	Each curve, corresponding to a different value of $g$, has a different optimal frequency: specifically, $\bar{\omega}$ decreases as the strength of reservoir connectivity $g$ increases from zero towards the edge-of-chaos (Fig.~\ref{fig:fig1}{\bf D}; see Appendix \ref{app:EOC} for a characterization of the edge-of-chaos in our framework).
	Although the exact value of the preferred frequency $\bar{\omega}$ is found to be algorithm-dependent, the same qualitative behaviour is observed consistently across the three different algorithms we used for training. It is also observed for both small and large amplitudes of the target signal $A$, which are expected to elicit, respectively,  weakly or strongly non-linear activity in the reservoir.
	
	The observations from Fig.~\ref{fig:fig1} provide a striking example of the non-trivial interplay between reservoir features (the connectivity parameter $g$) and external task parameters (the target frequency $\omega$) in determining learning performance. 
	Because the network is completely random, one might naively think that its dynamics do not exhibit a typical timescale, and are thus blind to the signal frequency; instead, the network appears to have its preference even for a simple task.
	In the rest of this paper, we aim to understand this observation in detail through mathematical analysis.
	
	To this end, we consider a simplified model which greatly eases the analysis: the case of linear reservoir dynamics ($\Phi(x)=x$).
	The analysis strategy we use consists of two steps \cite{RivkindBarak}. 
	To begin with, we examine the feedback network in an open-loop setup (Fig.~\ref{fig:fig1}{\bf A}, yellow), where the encoding of the input and the decoding of the output signals can be analyzed separately. In the encoding phase, we take the input to the reservoir network to be identical to the target function: $u(t)=f(t)$, and characterize analytically the reservoir response $\mathbf{x}(t)$ both at the level of single units and the population as a whole (Section \ref{sec:openloop_encoding}). 
	In the decoding phase, we use the reservoir response to pick a readout  $\mathbf{n}$ which allows the network to reconstruct the correct output: 
	$\mathbf{n}^\top\mathbf{x}(t)=f(t)$ (Section \ref{sec:openloop_decoding}).
	At that point, our feedback architecture admits the desired target as a solution; to investigate success of such solutions in performing the task, in Section \ref{sec:closed_loop} we close the loop (Fig.~\ref{fig:fig1}{\bf A}, purple), and characterize dynamics stability.
	Taken together, the open- and closed-loop descriptions fully characterize trained feedback architectures, and thus allow us to make predictions about dynamical mechanisms and training performance. 
	These predictions are shown to compare favorably to numerical simulations obtained by training linear feedback networks (Section \ref{sec:linear_performance}).
	Finally, in Section \ref{sec:non_linear}, we show that they qualitatively carry over to the case of non-linear dynamics.
	
	
	\subsection{Open loop: encoding the target signal}\label{sec:openloop_encoding}
	
	We begin our analysis by examining encoding in the open-loop framework: this corresponds to a random reservoir with linear dynamics driven by the target signal. The time evolution is described by
	\begin{equation}\label{eq:linear_ODE}
	\dot{\mathbf{x}}(t) = -\mathbf{x}(t) + \mathbf{J}\mathbf{x}(t) + \mathbf{m} f(t);
	\end{equation}
	here $\mathbf{J}$ is a Gaussian random matrix as defined above; to avoid dynamic instabilities, we consider $g<1$ \cite{Girko}. The linear dynamics is indifferent to the amplitude of the input, so we set $A=1$; in response to the periodic input $f(t) = \text{cos}(\omega t)$, the stationary solution for $t\to \infty$ is
	\begin{equation}\label{eq:x+-}
	\mathbf{x} (t) = \frac{1}{2} \left(\mathbf{x}_+ e^{i\omega t} +  \mathbf{x}_- e^{-i\omega t} \right),
	\end{equation}
	where
	\begin{equation}\label{eq:x+-2}
	\mathbf{x}_{\pm} := [(1\pm i\omega) \mathbf{I} - \mathbf{J}]^{-1} \mathbf{m}
	\end{equation}
	are complex conjugate vectors representing the reservoir activity in Fourier space (see Appendix \ref{app:linear_ODE}).
	
	\paragraph{A geometric description}
	
\begin{figure}
		
		\centering
		\includegraphics{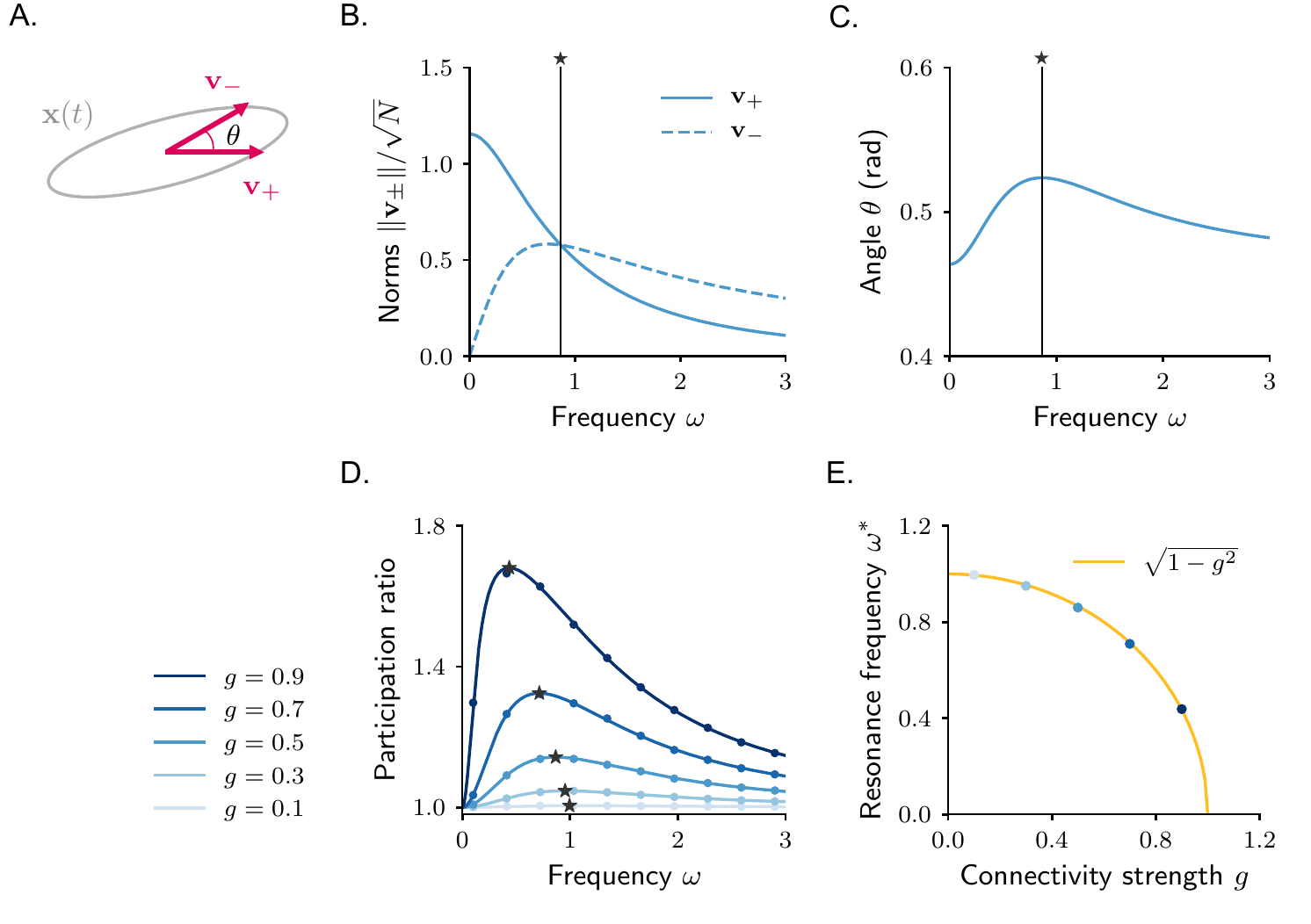}
		\caption{
			\textbf{Encoding of the target signal: geometry of activity trajectories.}
			{\bf A.} Projection of one example trajectory $\mathbf{x}(t)$ (Eq.~\eqref{eq:driven_state}) on the plane spanned by vectors $\mathbf{v}_\pm$.
			{\bf B.} Norm of the two vectors $\mathbf{v}_{\pm}$ as a function of the target frequency $\omega$.
			{\bf C.} Angle between the two spanning vectors $\mathbf{v}_\pm$.
			In {\bf A-B-C} we used $g=0.5$. In {\bf B-C}, the black vertical line indicates the resonance frequency $\omega^*$ where the two norms are equal ($r=1$, {\bf B}) and the angle is maximized ({\bf C}).
			{\bf D.} Linear dimensionality: participation ratio computed from the principal components of activity. We plot results for five increasing values of $g$ (blue shades); black stars indicate the position of $\omega^*$ for every value of $g$. 
			{\bf E.} Resonance frequency $\omega^*$.
			In all panels, continuous lines indicate the analytical predictions. In {\bf D}, dots show results averaged over $20$ simulations of finite size networks, $N=2000$.
			\label{fig:Encoding_Trajectories}
		}
	\end{figure}

	The stationary solution may be written as
	\begin{equation}\label{eq:driven_state}
	\mathbf{x}(t) = \mathbf{v}_+ \text{cos}(\omega t) + \mathbf{v}_- \text{sin}(\omega t),
	\end{equation}
	showing that activity occupies the plane spanned by two vectors $\mathbf{v}_{\pm}$, which are  the real and imaginary parts of $\mathbf{x}_{\pm}$: $\mathcal{R}(\mathbf{x}_{+})=\mathbf{v}_{+}$ and $\mathcal{I}(\mathbf{x}_{+})=-\mathbf{v}_{-}$. 
	In this plane, the state-space trajectory is a closed, elliptic curve
	(Fig.~\ref{fig:Encoding_Trajectories}{\bf A}), with
	geometry determined by the spanning vectors.

	The spanning vectors $\mathbf{v}_{\pm}$, in turn, depend both on the recurrent connectivity $\mathbf{J}$ and on the driving frequency $\omega$ (Eq.~\eqref{eq:x+-2}). Their geometry is self-averaging in the limit of large networks, and can be computed by averaging over the ensemble of randomly connected reservoir networks (see Appendix \ref{app:spanning_vecs}).
	Fig.~{\ref{fig:Encoding_Trajectories}{\bf B}} shows the dependence upon $\omega$ of the norms $\|\mathbf{v}_{\pm} \|$. 
	For very small frequencies, the trajectory follows the drive adiabatically and $\mathbf{v_-}\approx 0$; there is practically only one spanning vector. 
	As frequency increases, the response acquires a phase shift and the second spanning vector $\mathbf{v_-}$ becomes non-negligible. 
	At high frequencies, both norms decrease due to the filtering property of the network; the second spanning vector thus obtains a maximal norm at an intermediate frequency.
	
	We quantify the elliptical trajectory by its linear dimensionality, i.e.~the participation ratio computed from the principal components of reservoir activity \cite{GaoGanguli, LitwinKumar2017}.
	Denoting the activity cross-correlation matrix by $C := \frac{1}{T}\int_{0}^{T}\mathbf{x}(t)\mathbf{x}^\top\: \diff t$ and its eigenvalues by $\nu_i$, the trajectory dimensionality $d$ is defined as
	\begin{equation}
	d := \frac{\left(\sum_{i=1}^N \nu_i\right)^2}{\sum_{i=1}^N \nu_i^2}.
	\end{equation}
	By using Eq.~\eqref{eq:driven_state}, and by integrating out time, we find that $C$ is a rank-two matrix, $C = \frac{1}{2}\left(\mathbf{v}_+ \mathbf{v}_+^\top + \mathbf{v}_-\mathbf{v}_-^\top\right)$, whose non-zero eigenvalues (which we take to be $\nu_1, \nu_2$) are identical to those of the $2\times 2$ reduced cross-correlation matrix \cite{Nakatsukasa}
	\begin{equation} \label{eq:Amatrix}
	{C}^R = \begin{pmatrix}
	\|\mathbf{v}_+ \|^2      &  \mathbf{v}_+ \cdot \mathbf{v}_- \\
	\mathbf{v}_+ \cdot \mathbf{v}_- & \|\mathbf{v}_- \|^2
	\end{pmatrix}.
	\end{equation}
	Explicitly computing the eigenvalues of ${C}^R$ yields the expression
	\begin{equation} \label{eq:participation_ratio}
	d =\frac{\left(\|\mathbf{v}_+ \|^2 + \|\mathbf{v}_- \|^2\right)^2}{\|\mathbf{v}_+ \|^4 + 2\left(\mathbf{v}_+ \cdot \mathbf{v}_-\right)^2 + \|\mathbf{v}_- \|^4}.
	\end{equation}
	We observe that the linear dimensionality, which is bounded between 1 and 2, is insensitive to the overall trajectory magnitude, but depends on the ratio of norms $r=\|\mathbf{v}_- \|/\|\mathbf{v}_+ \|$ and on the angle $\theta$ between the spanning vectors:
	\begin{equation}
	d=\frac{1}{1-\frac{2r^2}{(1+r^2)^2}\,\sin^2(\theta)}.
	\end{equation}
	The ratio $r$ indicates how much the curve is squeezed along a single direction, with both extremes ($r$ very small or very large) resulting in trajectories squeezed along the dominant spanning vector. 
	For a fixed angle, as the ratio passes through $r=1$, the trajectory goes through a shape which is most similar to a circle and has maximal dimensionality. 
	For a fixed $r$, the angle $\theta$ determines to what degree the curve is skewed relative to a perfect ellipse; the dimensionality increases monotonically as $\theta$ opens up from zero to $\pi/2$. 
	
	Examination of the vector norms $\|\mathbf{v}_{\pm} \|$ in Fig.~\ref{fig:Encoding_Trajectories}{\bf B} indicates that they intersect at a frequency value that we name $\omega^*$, where $r=1$.
	Fig.~\ref{fig:Encoding_Trajectories}{\bf C} shows how the angle $\theta$ varies as a function of frequency; surprisingly, we find that it displays a maximum at $\omega^*$. 
	These dependencies are reflected in the behaviour of the dimensionality (Fig.~\ref{fig:Encoding_Trajectories}{\bf D}), which itself attains a maximum at frequency $\omega^*$. 
	Our mathematical analysis reveals that
	(see Appendix \ref{app:geometric_analysis})
	\begin{equation}
	\omega^*=\sqrt{1-g^2},
	\end{equation}
	i.e.~the resonance frequency $\omega^*$  decreases to zero as $g$ increases towards the instability boundary ($g=1$). 
	This analytic result is in excellent agreement with finite network simulations, as shown in Fig.~\ref{fig:Encoding_Trajectories}{\bf E}.

	\paragraph{A single-unit description}
	
	The analysis above considered the geometry of trajectories spanned by the reservoir population in its high-dimensional activity space, and revealed that trajectory dimensionality is maximized at the resonance frequency $\omega^*$. 
	An alternative viewpoint is obtained by considering the statistics of single-unit activity profiles across the population. 
	As we shall see, this alternative perspective reveals that the optimal frequency $\omega^*$ has a second natural interpretation in terms of population synchrony.
	
	To do so, we derive a self-consistent expression for $\mathbf{x}_+$ by inserting Eq.~\eqref{eq:x+-} into the evolution equations (Eq.~\eqref{eq:linear_ODE}):
	\begin{equation}\label{eq:x+_d}
	\mathbf{x}_+ = \frac{1}{1+i\omega} \left( {\mathbf{m}} + \mathbf{J} \mathbf{x}_+   \right).
	\end{equation}
	This form highlights that vector $\mathbf{x}_+$ is given by the sum of two contributions: one associated with the external forcing  via the input vector $\mathbf{m}$, and one associated with the reservoir response via the  recurrent input $\mathbf{Jx}_+$. Since $\mathbf{J}$ is random, the direction of the latter contribution is random (i.e., it varies across realizations of $\mathbf{J}$), but its amplitude is self-averaging and depends on the strength of recurrent connectivity $g$ \cite{Sompolinsky1988}.
	
\begin{figure}
		
		\centering
		\includegraphics{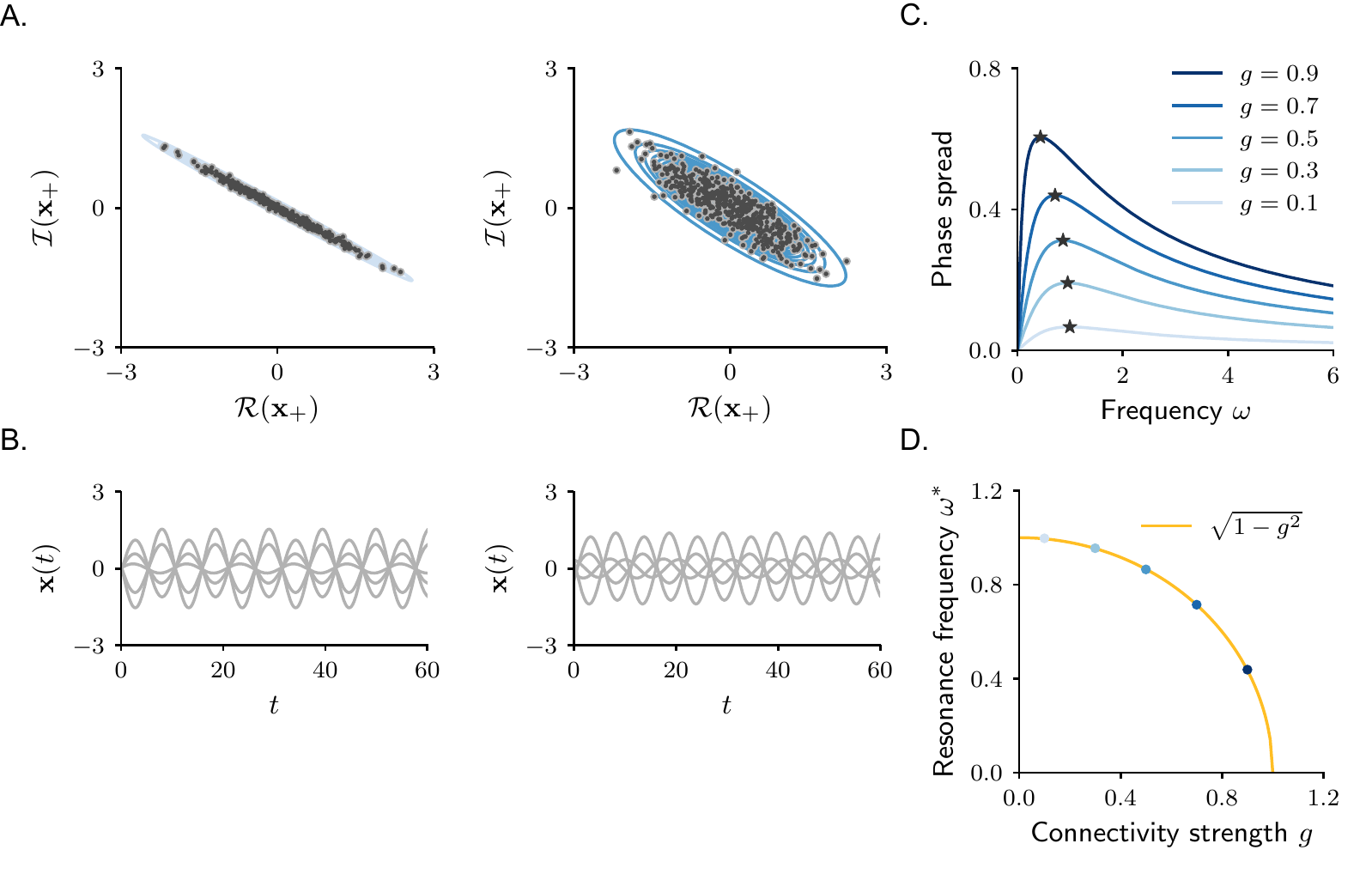}
		\caption{ {\bfseries Encoding of the target signal: single-unit description.} 
			{\bfseries A.} Entries of vector $\mathbf{x}_+$ in the complex plane. Left: $g=0.1$, right: $g=0.5$; $w=0.6$ for both panels. 
			Continuous lines are contour lines of the bivariate Gaussian distribution predicted by the theory. The most external contour indicates a probability of 0.01. 
			Grey points are results of a finite network simulation ($N=400$).
			{\bfseries B.} Sample of activity from four randomly selected units chosen from the corresponding panel in {\bf A}.
			{\bfseries C.} Spread of response phases across the population (Eq.~\eqref{eq:phasespread}) for increasing values of $g$ and as a function of $\omega$. Black stars indicate the maximum value.
			{\bfseries D.} Value of the frequency which maximizes the spread of phases from {\bfseries C}.
		}
		\label{fig:Encoding_SingleUnits}
		
	\end{figure}

	We use Eq.~\eqref{eq:x+_d} to gain intuition about how the network encodes the external oscillatory signal at the level of single-unit activity. To this end, we visualize the entries of the $\mathbf{x}_+$ vector as points in the complex plane: $(\mathbf{x}_+)_i={R_i}e^{i{\phi_i}}$, where $R_i$ and $\phi_i$ represent the amplitude and phase with which a single unit responds to the forcing input (Fig.~\ref{fig:Encoding_SingleUnits}{\bf A}). 
	How are points corresponding to different units distributed on the complex plane? When recurrent connections are very weak ($g\simeq0$),  different units behave as uncoupled filters of the input; we have $\mathbf{x}_+ \simeq {\mathbf{m}}/(1+i\omega)$, implying that the real and imaginary part of $(\mathbf{x}_+)_i$ for different $i$ are proportional one to each other. 
	As a consequence, points on the complex plane are collinear (Fig.~\ref{fig:Encoding_SingleUnits}{\bf A} left), and phases are identical: $\phi_i=\phi$. Responses of different units are thus synchronized (Fig.~\ref{fig:Encoding_SingleUnits}{\bf B} left). As $g$ grows from 0, the second term in Eq.~\eqref{eq:x+_d}, which originates from recurrent interactions, starts spreading the real and imaginary parts of $(\mathbf{x}_+)_i$ away from the line $\phi_i=\phi$ (Fig.~\ref{fig:Encoding_SingleUnits}{\bf A} right), and introduces variability in response phases (Fig.~\ref{fig:Encoding_SingleUnits}{\bf B} right).
	
	For fixed values of $g$ and $\omega$, the distribution of dots on the complex plane is a bivariate Gaussian (Fig.~\ref{fig:Encoding_SingleUnits}{\bf A}); a \emph{narrow} distribution corresponds to highly 
	synchronized units, and its \emph{broadening} at stronger coupling indicates their desynchronization.
	As both $\mathbf{m}$ and $\mathbf{J}$ are generated from a centered Gaussian distribution, the mean of the distribution vanishes.
	The covariance is given by:
	\begin{equation}
	\frac{1}{N}
	\begin{pmatrix}
	\|\mathbf{v}_+\|^2& - \mathbf{v}_+ \cdot \mathbf{v}_- \\
	- \mathbf{v}_+ \cdot \mathbf{v}_- & \|\mathbf{v}_-\|^2
	\end{pmatrix}
	\end{equation}
	implying that the shape distribution is controlled by the statistics of the spanning vectors $\mathbf{v}_+$ and $\mathbf{v}_-$.
	The similarity between the covariance matrix and the reduced cross-correlation matrix $C^R$ (Eq.~\eqref{eq:Amatrix}) analyzed in the previous paragraph suggests that synchrony in single-unit response and dimensionality of state-space trajectories are deeply related properties of reservoir activity.
	To formalize this relationship, we compute the spread of phases $\phi_i$ across the reservoir population, i.e.
	\begin{equation}\label{eq:phasespread}
	\Sigma^2 = \int_{\bar{\phi}-\frac{\pi}{2}}^{\bar{\phi}+\frac{\pi}{2}} \diff \phi \: p(\phi) (\phi-\bar{\phi})^2,
	\end{equation}
	where $p(\phi)$ is the probability distribution of phases for a bivariate Gaussian distribution (\cite{Aalo2007}, see Appendix \ref{app:phases}).
	The phase spread for different values of recurrent strength $g$ and frequency $\omega$ is plotted in Fig.~\ref{fig:Encoding_SingleUnits}{\bf C}. These results show that it monotonically increases with $g$; for any fixed $g$, it reaches a maximum at a finite frequency value, given again by $\omega^*=\sqrt{1-g^2}$ (Fig.~\ref{fig:Encoding_SingleUnits}{\bf D}).
	
	To conclude, we have examined the behaviour of single-unit activity in response to a sinusoidal forcing input. 
	In line with classical mean-field studies, we have analyzed the statistical distribution of single-unit activity profiles across the reservoir population \cite{Sompolinsky1988, Rajan2010, MassarMassar}. 
	This approach has revealed that, for fixed $g$, $\omega^*$ corresponds to the frequency at which single-unit activity is maximally desynchronized.
	Note that historically, desynchronized single-unit profiles have been pointed out as a desirable feature of reservoir activity, as temporally heterogeneous profiles form a rich set of basis functions from which complex target functions can be reconstructed \cite{Maass2002}.

	\subsection{Open-loop setup: decoding the internal representation}
	\label{sec:openloop_decoding}
	
	After having characterized the reservoir activity during stimulus encoding, we turn to the decoding step of the open-loop analysis. Decoding corresponds to finding a readout vector $\mathbf{n} \in \mathbb{R}^N$ which satisfies:
	\begin{equation} \label{eq:readout-eq}
	z(t) = \mathbf{n}^\top\mathbf{x}(t) = \cos(\omega t);
	\end{equation}
	the projection of driven reservoir activity along $\mathbf{n}$ needs thus to match the target $f(t)$  (Fig.~\ref{fig:ClosedLoop}{\bf A}, yellow).
	In terms of the Fourier-space representation (Eq.~\eqref{eq:x+-}), $\mathbf{n}$ is a solution to the set of two linear equations given by
	\begin{equation} \label{eq:decoding_fourier}
	\mathbf{n}^\top \mathbf{x}_\pm = \mathbf{n}^\top[(1\pm i\omega)\mathbf{I} - \mathbf{J}]^{-1}\mathbf{m} = 1. 
	\end{equation}
	When $g=0$, interactions vanish and the equations above read $\mathbf{n}^\top \mathbf{m} = 1 \pm i \omega$, which cannot be satisfied by any $\mathbf{n}$. 
	This scenario corresponds to completely synchronized reservoir activity, or equivalently, activity spanning one-dimensional state-space trajectories. 
	For any $g>0$, on the other hand, this system of equations is under-determined, since it fixes only 2 among the $N$ degrees of freedom in $\mathbf{n}$. 
	
	We explore the effect of these degrees of freedom by defining a family of readout vectors $\mathbf{n}$ parametrized by an integer $k$, where $k=2,\ldots,N$; $k$ indicates the number of reservoir units from which the readout signal is reconstructed.
	We term such solutions \emph{from-}$k$ regression.
	To obtain such a solution, we set  all elements except for the first $k$ of $\mathbf{n}$ to zero, and then solve Eq.~\eqref{eq:decoding_fourier} by considering the least-squares (LS) solution of minimal norm, which can be computed through the pseudo-inverse (see Appendix \ref{app:from-k}). 
	When $k=N$, we obtain the full LS solution, which reads:
	\begin{equation}\label{eq:readoutx}
	\mathbf{n}_{LS} = 
	\begin{pmatrix}
	\mathbf{x}_+ & \mathbf{x}_-
	\end{pmatrix}
	{\begin{pmatrix}
		\| \mathbf{x}_+ \|^2 & \mathbf{x}_+ \cdot \mathbf{x}_- \\
		\mathbf{x}_+ \cdot \mathbf{x}_- & \| \mathbf{x}_- \|^2 
		\end{pmatrix}}^{-1}
	\begin{pmatrix} 1 \\ 1 \end{pmatrix}
	\end{equation}
	or, in terms of $\mathbf{v_\pm}$ vectors:
	\begin{equation}\label{eq:readoutv}
	\begin{aligned}
	\mathbf{n}_{LS} = 
	\begin{pmatrix}
	\mathbf{v}_+ & \mathbf{v}_-
	\end{pmatrix}
	P
	\begin{pmatrix} 1 \\ 0 \end{pmatrix},
	\end{aligned}
	\end{equation}
	where we defined the short-hand notation $P=\left(C^R\right)^{-1}$.
	
	All the readouts within the \emph{from-k} family exactly solve the task in the open-loop setup. However, it is not clear \textit{a-priori} whether all of them are equivalent when closing the loop, i.e.~when the feedback network is required to autonomously generate the target signal (Eq.~\eqref{eq:dyn_feedback}).
	In the following, we assess dynamics and stability of closed-loop networks corresponding to the different choices of the readout $\mathbf{n}$.
	
	\subsection{Closing the loop: autonomous signal generation} \label{sec:closed_loop}
	
\begin{figure}
		
		\centering
		\includegraphics{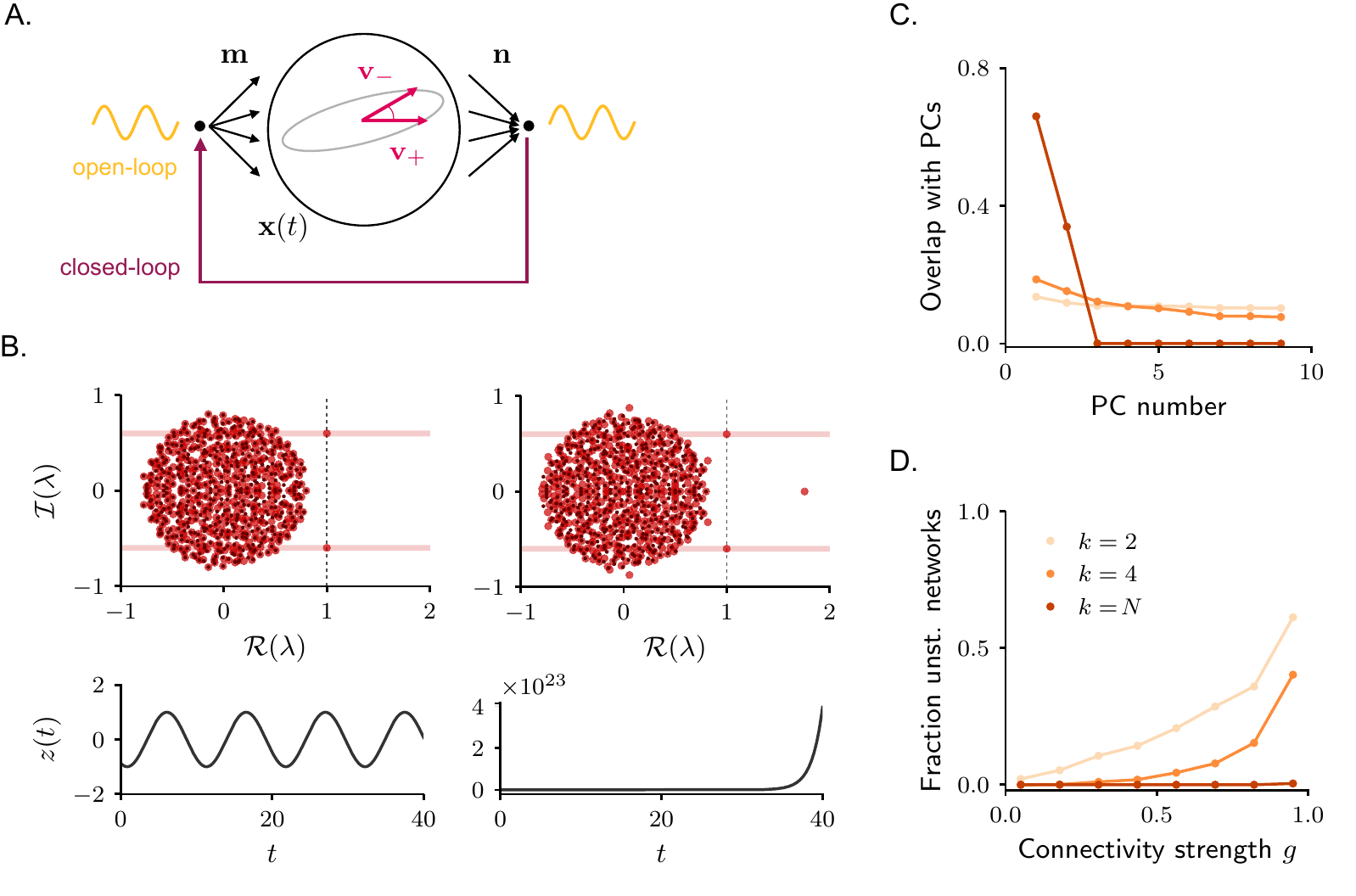}
		
		\caption{ {\bfseries Closing the loop.} 
			{\bfseries A.} Transforming the open-loop encoding/decoding setup (yellow) into a closed-loop system (purple). 
			{\bfseries B.} Sample networks trained through full LS (\emph{from-N}, left) or \emph{from-2} (right) regression.
			The top panels show the eigenspectra of the closed-loop connectivity matrix $\mathbf{\bar{J}}$ (red dots); small black dots indicate the unperturbed eigenspectrum of $\mathbf{{J}}$. 
			The bottom panels show the output generated by the corresponding networks.
			Here we used parameters $g=0.8$, $\omega=0.6$ and $N=400$.
			{\bfseries C.} Overlap between readout $\mathbf{n}$ and the principal components (PC) of driven reservoir activity (Eq.~\eqref{eq:driven_state}), of which the first two span the $\mathbf{v}_\pm$ plane.
			The same parameters as in {\bfseries B} were used.
			{\bfseries D.} Fraction of unstable closed-loop systems (over 2000 sample networks) as a function of connectivity strength $g$, for several values of $k$, measured over 1000 different realizations.
			We used $N=1000$ and $\omega=0.6$.
		}
		\label{fig:ClosedLoop}
		
	\end{figure}

	In the previous two sections, we have analyzed how random networks encode a one-dimensional periodic signal, and how the network response can be used to reconstruct the same signal as output. 
	Ultimately, we want the encoding and the decoding steps to be self-consistent, i.e.~we require
	\begin{equation}
	z(t)= \mathbf{n}^\top \mathbf{x} (t) = u(t) = f(t),
	\end{equation}
	which is equivalent to transforming our problem from an open-loop to a closed-loop setup, where dynamics are autonomous and follow Eq.~\eqref{eq:dyn_feedback} with linear interactions: 
	\begin{equation}\label{eq:dyn_feedback_lin}
	\dot{\mathbf{x}}(t) = -\mathbf{x}(t) + (\mathbf{J} + \mathbf{m} \mathbf{n}^\top) \mathbf{x}(t)
	\end{equation}
	and $\mathbf{n}$ satisfies Eq.~\eqref{eq:readout-eq}. 
	This step is illustrated in Fig.~\ref{fig:ClosedLoop}{\bf A} by the purple feedback arrow connecting the reservoir output to the input. If closing the loop does not perturb activity in the reservoir by changing its stability properties, then at every time point the readout $\mathbf{n}^\top \mathbf{x}(t)=\cos(\omega t)$ is fed back into the system, and the solution obtained through the open-loop setup is self-consistent.
	
	The solutions to Eq.~\eqref{eq:dyn_feedback_lin} and their stability are fully characterized by the eigenspectrum of $\mathbf{\bar{J}} = \mathbf{J} + \mathbf{m}\mathbf{n}^\top$ (the leak term in the dynamics contributes by uniformly shifting the spectrum by $-1$).
	For $N$ sufficiently large, the eigenvalues of $\mathbf{J}$ are distributed uniformly in a disk of radius $g<1$ \cite{Girko}. 
	The position of some or all of the eigenvalues can, however, be modified by the rank-one perturbation $\mathbf{mn}^\top$; we refer to these as \emph{outliers}.
	In order for the closed-loop system to stably sustain the periodic activity we found in the encoding step, the eigenspectrum of  $\mathbf{\bar{J}}$ must satisfy two key requirements: (i) a pair of complex outlier eigenvalues with value: $\lambda_{\pm} = 1 \pm i \omega$ (which ensures that a periodic trajectory of frequency $\omega$ is realized), and (ii) a stable bulk of remaining eigenvalues: $\mathcal{R}(\lambda)<1$ $\forall \lambda \neq \lambda_\pm$ (which ensures that no runaway activity is generated along other directions).
	
	All eigenvalues of $\mathbf{\bar{J}}$ are roots of the characteristic polynomial
	\begin{equation}
	\det\left( (\mathbf{J} + \mathbf{m}\mathbf{n}^\top) - \lambda \mathbf{I} \right)= 0.
	\end{equation}
	The Matrix Determinant Lemma \cite{Logiaco2019, Schuessler2020} allows us to decompose this polynomial into two factors, corresponding to the two sets of eigenvalues:
	\begin{equation}
	\det\left( 1 + \mathbf{n}^\top(\mathbf{J}-\lambda\mathbf{I})^{-1}\mathbf{m} \right) \det \left( \mathbf{J} - \lambda\mathbf{I} \right)= 0.
	\end{equation}
	It is seen that the second term vanishes on the spectrum of $\mathbf{J}$, whereas the first term vanishes for the outlier eigenvalues. 
	The outliers therefore satisfy
	\begin{equation}\label{eq:MDL}
	1 = \mathbf{n}^\top(\lambda\mathbf{I}-\mathbf{J})^{-1}\mathbf{m}.
	\end{equation}
	If $\mathbf{n}$ satisfies Eq.~\eqref{eq:readout-eq}, then  $\lambda_\pm = 1 \pm i\omega$ are indeed solutions, implying that condition (i) is satisfied. 
	Note that the eigenvectors corresponding to $\lambda_{\pm}$ are identical to the vectors $\textbf{x}_\pm$, as $(\mathbf{J} + \mathbf{m}\mathbf{n}^\top) \mathbf{u}_\pm = \lambda_\pm \mathbf{u}_\pm$
	implies
	\begin{equation} \label{eq:CLeigenvecs}
	\mathbf{u}_\pm \propto (\mathbf{J} - \lambda_\pm \mathbf{I})^{-1} \mathbf{m}.
	\end{equation}
	Thus fixing $\mathbf{n}$ in the open-loop framework is equivalent to directly controlling the value of the target-relevant eigenvalues in the eigenspectrum of the closed-loop network. 
	
	We next examine whether this pair of eigenvalues are the only outliers generated by closing the loop: while Eq.~\eqref{eq:MDL} is guaranteed to have $\lambda_{\pm}=1\pm i\omega$ as solutions, other solutions might be admitted which could violate requirement (ii).
	Such potential solutions depend on the overlap between the vectors $\mathbf{n}$ and $\mathbf{x}_{\lambda} = (\lambda\mathbf{I}-\mathbf{J})^{-1}\mathbf{m}$.
	Note that if the readout $\mathbf{n}$ was random (and thus orthogonal to $\mathbf{J}$ and $\mathbf{m}$), this overlap would vanish and no additional outliers would be generated.
	
	In the case of the full LS solution ($k=N$, Eqs.~\eqref{eq:readoutx}-\eqref{eq:readoutv}), the readout vector  $\mathbf{n}_{LS}$ is contained in the plane spanned by vectors $\mathbf{v}_\pm$. As a consequence, the overlap between $\mathbf{n}$ and $\mathbf{x}_{\lambda}$ can be expanded in terms of $\mathbf{x}_\pm^\top \mathbf{x}_{\lambda}$. 
	In the limit $N\rightarrow\infty$, these terms have a simple form which can be evaluated analytically (see Appendix \ref{app:LS_eigs}), yielding an equation in $\lambda$ which reads:
	\begin{equation}\label{eq:quadratic_eq}
	(1+\omega^2){\lambda}^2 - \left[ 2g^2 + N (P_{11} + \omega P_{21}) \right] {\lambda}  +  g^4 + Ng^2P_{11} = 0
	\end{equation}
	where $P_{11}$ and $P_{21}$ are the elements of the first column of $P=\left(C^R\right)^{-1}$ and depend on $g$ and $\omega$.
	As the equation above is quadratic, it admits $\lambda=\lambda_\pm$ as unique solutions.
	Therefore, in large networks, LS training is guaranteed to result in stable dynamics, as no additional outliers are generated in the eigenspectrum other than the task-relevant ones. 
	This is confirmed by numerical simulation in the left panels of Fig.~\ref{fig:ClosedLoop}{\bf B}. 
	
	In the more general case of \emph{from-k} LS regressors with $k<N$, readout vectors might contain extra components that are correlated with $\mathbf{J}$ and $\mathbf{m}$ and are not fully contained within the spanning plane; as a consequence, more than two outlier eigenvalues and unstable dynamics can be expected.  
	The right panels of Fig.~\ref{fig:ClosedLoop}{\bf B} show an example of such a situation (in the simulation, $k=2$). 
	One outlier eigenvalue with $\mathcal{R}(\lambda)>1$ is seen in the top panel, which induces the dynamic instability seen in the bottom panel.
	
	As $k$ decreases from the maximal value ($N$) to the minimal one ($2$), the component of the readout vector $\mathbf{n}$ outside of the $\mathbf{v}_\pm$ plane becomes larger (Fig.~\ref{fig:ClosedLoop}{\bf C}). 
	Numerical analysis indicates that, correspondingly, the fraction of networks with unstable dynamics increases (Fig.~\ref{fig:ClosedLoop}{\bf D}).
	
	In summary, we have shown that -- although the open-loop setup admits multiple exact solutions --  different solutions are not equivalent in terms of dynamical stability when the loop is closed. 
	Stability properties are related to the orientation of the readout vector relative to the driven open-loop trajectory.
	In the case of the full LS solution (Eq.~\eqref{eq:readoutx}), 
	the readout $\mathbf{n}_{LS}$ is completely aligned with the trajectory plane, and closed-loop dynamics are guaranteed to be stable.
	Other solutions generally contain components outside of this plane, which can cause activity to diverge.

	\subsection{Predicting performance of trained linear networks}\label{sec:linear_performance}
	
	We now turn back to the problem of understanding performance in trained feedback networks and its dependence on the target frequency $\omega$. 
	We start by considering linear feedback networks which are trained (as in Fig.~\ref{fig:fig1}{\bf C} left) via LS regression.
	The analysis of the previous sections has shown that linear networks trained via LS regression can exactly implement the feedback task with stable dynamics.
	Due to noise, however, real-life LS regression never converges to this ideal solution. 
	Noise arises in training from multiple sources, such as finite sampling of training data, variability due to different initial conditions or regularization noise. 
	In order to characterize training performance, we thus use our theoretical framework to  analyze the effect of noise on the dynamics of feedback linear networks trained via LS regression.
	
	Consider first the encoding phase of learning (Section \ref{sec:openloop_encoding}), where reservoir activity is stimulated. Because of noise, learning algorithms may not have access to the true spanning vectors $\mathbf{v}_\pm$. Rather, we assume that corrupted versions $\tilde{\mathbf{v}}_\pm = \mathbf{v}_\pm + \bm{\xi}_\pm$ (where the entries of $\bm{\xi}_\pm$ are independent Gaussian noise) are measured.
	The estimated LS readout then reads:
	\begin{equation}\label{eq:ntilde}
	\tilde{\mathbf{n}}_{LS} = 
	\begin{pmatrix}
	\mathbf{v}_+ & \mathbf{v}_-
	\end{pmatrix}
	\tilde{P}
	\begin{pmatrix} 1 \\ 0 \end{pmatrix}
	\: + \: \begin{pmatrix}
	\bm{\xi}_+ & \bm{\xi}_-
	\end{pmatrix}
	\tilde{P}
	\begin{pmatrix} 1 \\ 0 \end{pmatrix}
	\end{equation}
	where $\tilde{P}$ is the inverse reduced cross-correlation matrix which includes the noise disturbance.
	
	As in Section \ref{sec:closed_loop}, we can characterize closed-loop dynamics by computing the outlier eigenvalues of $\mathbf{J}+\mathbf{m} \:{\tilde{\mathbf{n}}_{LS}}^\top$.
	The second term in the r.h.s.~of Eq.~\eqref{eq:ntilde} is random and orthogonal to $\mathbf{m}$ and $\mathbf{J}$, and therefore does not affect the position of outlier eigenvalues.
	In contrast, the first term is a vector fully aligned with the noise-free spanning vectors $\mathbf{v}_\pm$, which generates two outlier eigenvalues $\tilde{\lambda}_\pm$. 
	Because of the noise, their values deviate from the target eigenvalues $\lambda_\pm$; they are solutions of an equation identical to Eq.~\eqref{eq:quadratic_eq}, but with $P$ replaced by $\tilde{P}$.
	For every noise realization, the inverse reduced cross-correlation matrix $P$ is perturbed in a random direction, yielding random modifications to the target eigenvalues ${\lambda}_\pm$. 
	We can estimate the average mismatch between $\tilde{\lambda}_\pm$ and ${\lambda}_\pm$ from the sensitivity of matrix $P$ to perturbations, which is quantified by the condition number of the reduced correlation matrix $C^R$, i.e.
	the ratio between the largest and smallest eigenvalue \cite{Lukosevicius, JaegerRiddle}:
	\begin{equation}
	c = \frac{\nu_2}{\nu_1}.
	\end{equation}
	
\begin{figure}
		
		\centering
		\includegraphics{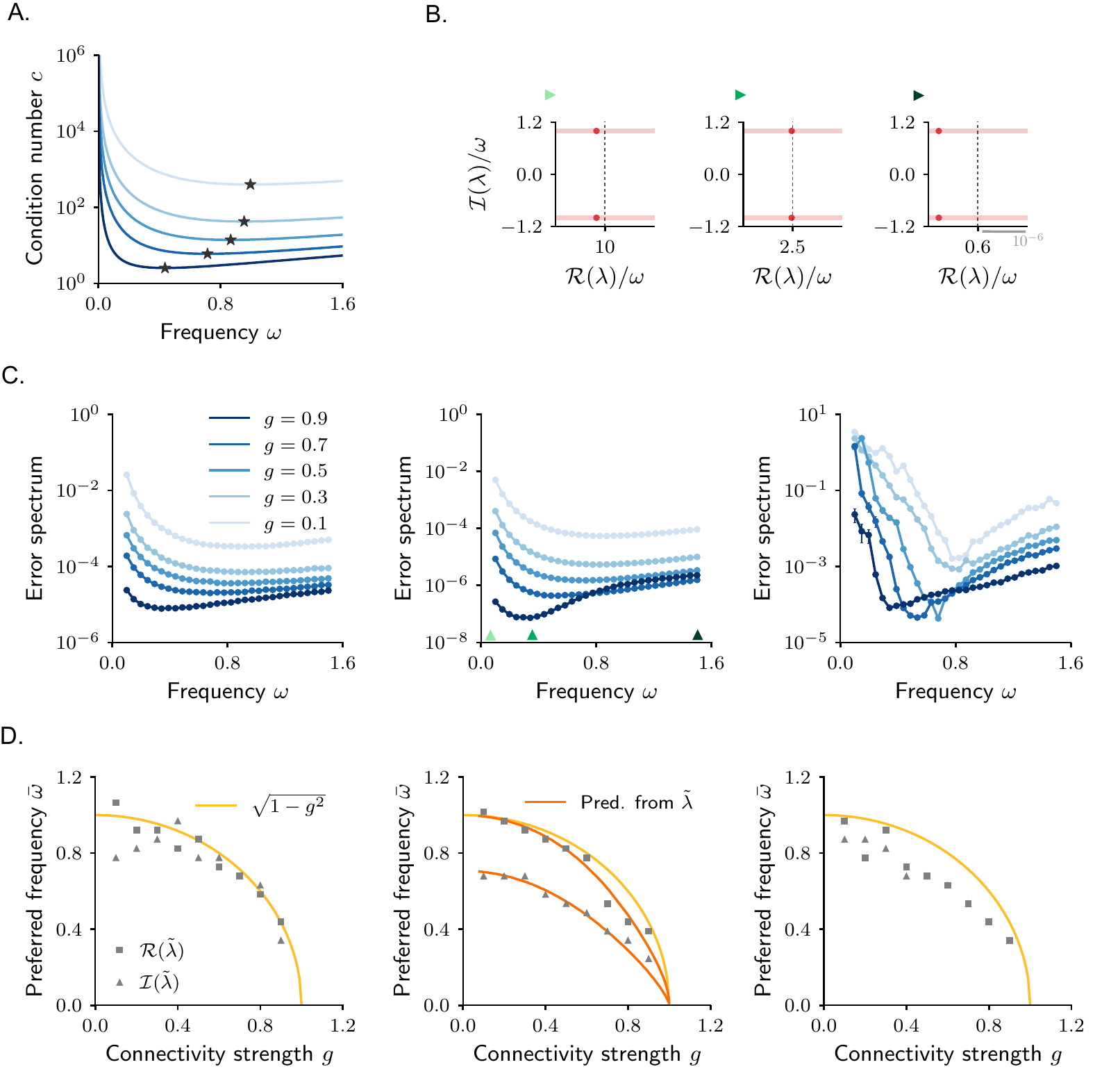}
		\caption{
			\textbf{Training performance in linear networks is maximized at $\omega^*$.}
			{\bf A.} Condition number of the reduced cross-correlation matrix $C^R$ computed analytically for different values of $g$ (blue shades in {\bf C}). Stars denote minimal condition number.
			{\bf B.} Closed-loop outlier eigenvalues $\tilde{\lambda}_\pm$ for example networks from three learning trials corresponding to the  frequencies marked by green triangles in {\bf C} ($w=$ 0.1, 0.4, 1.5; $g=0.9$). 
			{\bf C.} Spectrum error as a function of $\omega$, for a range of $g$ values (blue shades), for networks trained via noisy LS (left), ridge regression (middle) and RLS (right). 
			Error is measured from the imaginary part of outlier eigenvalues as $|\mathcal{I}(\tilde{\lambda}_\pm)-\mathcal{I}(\lambda_\pm)|/\omega$, and similarly for the real part; the spectrum error is an average of the two.
			Training details and parameters are reported in Appendix \ref{app:training}.
			{\bf D.} Frequency $\bar{\omega}$ minimizing the error in the real and imaginary part of $\tilde{\lambda}_\pm$ as a function of $g$, for the three training algorithms.
			Solid orange lines in the middle panel show theoretical prediction for ridge regression (see Appendix \ref{app:ridge_regression}).
			\label{fig:predicting_perf}
		}
	\end{figure}

	The value of $c$ and its dependence on $\omega$ and $g$ can be computed by taking the limit $N\rightarrow\infty$ and averaging over the networks ensemble.
	Fig.~\ref{fig:predicting_perf}{\bf A} shows that, for fixed connectivity strength $g$, the condition number is a non-monotonic function of the forcing frequency $\omega$, and attains a minimum at the resonance frequency $\omega^* = \sqrt{1-g^2}$ (see Appendix \ref{app:PR_CN_extrema}).
	Thus, when training linear feedback networks through noisy LS regression we expect that, for fixed $g$, the readout would be closest to the desired one at $\bar{\omega} = \omega^*$, where the reduced cross-correlation matrix is most robust to noise.
	This robustness directly reflects the properties of the internal representation of the target signal within the reservoir, which is characterized by maximal dimensionality and minimal synchrony at $\omega=\omega^*$. 
	
	We tested this prediction on finite-size trained networks.
	Examples from Fig.~\ref{fig:predicting_perf}{\bf B} confirm that the task-related eigenvalue pair $\tilde{\lambda}_\pm$ deviate from the target ones. 
	As in the case of non-linear networks (Fig.~\ref{fig:fig1}{\bf B-C}), we find that the error is frequency dependent (Fig.~\ref{fig:predicting_perf}{\bf C}, left); furthermore, for fixed strength of the internal connectivity $g$, we observe that the error is minimized at a frequency $\bar{\omega}$ which is very close to $\omega^*$ (Fig.~\ref{fig:predicting_perf}{\bf D}, left).

	As a second way to characterize performance, like in Fig.~\ref{fig:fig1}, we considered feedback networks trained via Ridge regression \cite{HoerlRidge}. 
	In this case, the readout vector is deterministic; in the Fourier space, it can be expressed as (see Appendix \ref{app:ridge_regression}):
	\begin{equation}
	\tilde{\mathbf{n}}_{R} = 
	\begin{pmatrix}
	\mathbf{v}_+ & \mathbf{v}_-
	\end{pmatrix}
	\tilde{P}
	\begin{pmatrix} 1 \\ 0 \end{pmatrix}
	\end{equation}
	where $\tilde{P} = \left(C^R\right)^{-1}+N\sigma^2 \mathbf{I}$ and $\mathbf{I}$ is the $2\times 2$ identity matrix. 
	As in the case of noisy LS regression, also this readout vector generates only two outlier eigenvalues $\tilde{\lambda}_\pm$, whose values can be again computed through Eq.~\eqref{eq:quadratic_eq}; in this case a closed-form expression can be computed (Appendix \ref{app:ridge_regression}).
	
	For moderate values of the regularization parameter $\sigma$, the resulting $\tilde{\lambda}_\pm$ are complex conjugates that  deviate somewhat from ${\lambda}_\pm$ (see Supp.~Fig.~\ref{fig:bifurcation_diagram} for the full bifurcation diagram). 
	Specifically, their real part is always smaller than 1, implying that the resulting autonomous dynamics are always stable (see Appendix \ref{app:ridge_regression}).
	The amplitude of the mismatch between the real and the imaginary parts of $\tilde{\lambda}_\pm$ and the target eigenvalues ${\lambda}_\pm$ depends both on $g$ and $\omega$ (see Appendix \ref{app:ridge_regression}), and is minimized at a finite frequency $\bar{\omega}$ which monotonically decreases with increasing $g$ (Fig.~\ref{fig:predicting_perf}{\bf D} center, solid lines).
	Importantly, the value of $\bar{\omega}$ is predicted to behave similarly (although not identically) to $\omega^*$. Fig.~\ref{fig:predicting_perf}{\bf D} (middle) shows an excellent match between these predictions and simulation results.
	
	As a third and final example, we considered linear networks trained via the RLS algorithm \cite{Liu, SussilloAbbott}.
	In this case, an analytical description of the closed-loop spectrum and resulting dynamics is much harder to obtain; we thus computed the value of the preferred frequency $\bar{\omega}$ from simulations.
	We found that the mismatch between $\tilde{\lambda}_\pm$ and ${\lambda}_\pm$ displays a strong, non-monotonic dependence on the target frequency (Fig.~\ref{fig:predicting_perf}{\bf C}, right); the preferred frequency $\bar{\omega}$ is, again, quite close to $\omega^*$ (Fig.~\ref{fig:predicting_perf}{\bf D}, right).
	
	To conclude, we analyzed performance in linear feedback networks; as for non-linear networks (Fig.~\ref{fig:fig1}), we found that performance is maximized for a preferred frequency $\bar{\omega}$ which decreases with the connectivity strength $g$. 
	Analysing how the simple LS readout solution interacts with noise, we predicted that the preferred frequency $\bar{\omega}$ is expected to lay close to $\omega^*$, i.e.~the resonance frequency where encoding dynamics has maximal dimensionality and is minimally synchronized.
	This prediction is exactly verified in networks trained via LS regression, but also carry over in a qualitative fashion to networks trained via different training algorithms. 
	In fact, we showed that different algorithms are affected by different kinds of biases, whose effect is to shift the value of the preferred frequency $\bar{\omega}$ away from $\omega^*$ without changing its overall qualitative behaviour.

	\subsection{Internal representation in non-linear networks}\label{sec:non_linear}
	
	We finally turn back to the original problem of analyzing training performance in non-linear feedback networks (Fig.~\ref{fig:fig1}).
	Our analysis of linear networks revealed that a key feature which determines training performance is the quality of representation of the target signal within the reservoir. 
	This representation can be characterised by its dimensionality or, equivalently, by the synchrony of activity across units in the reservoir. 
	
\begin{figure}
		
		\centering
		\includegraphics{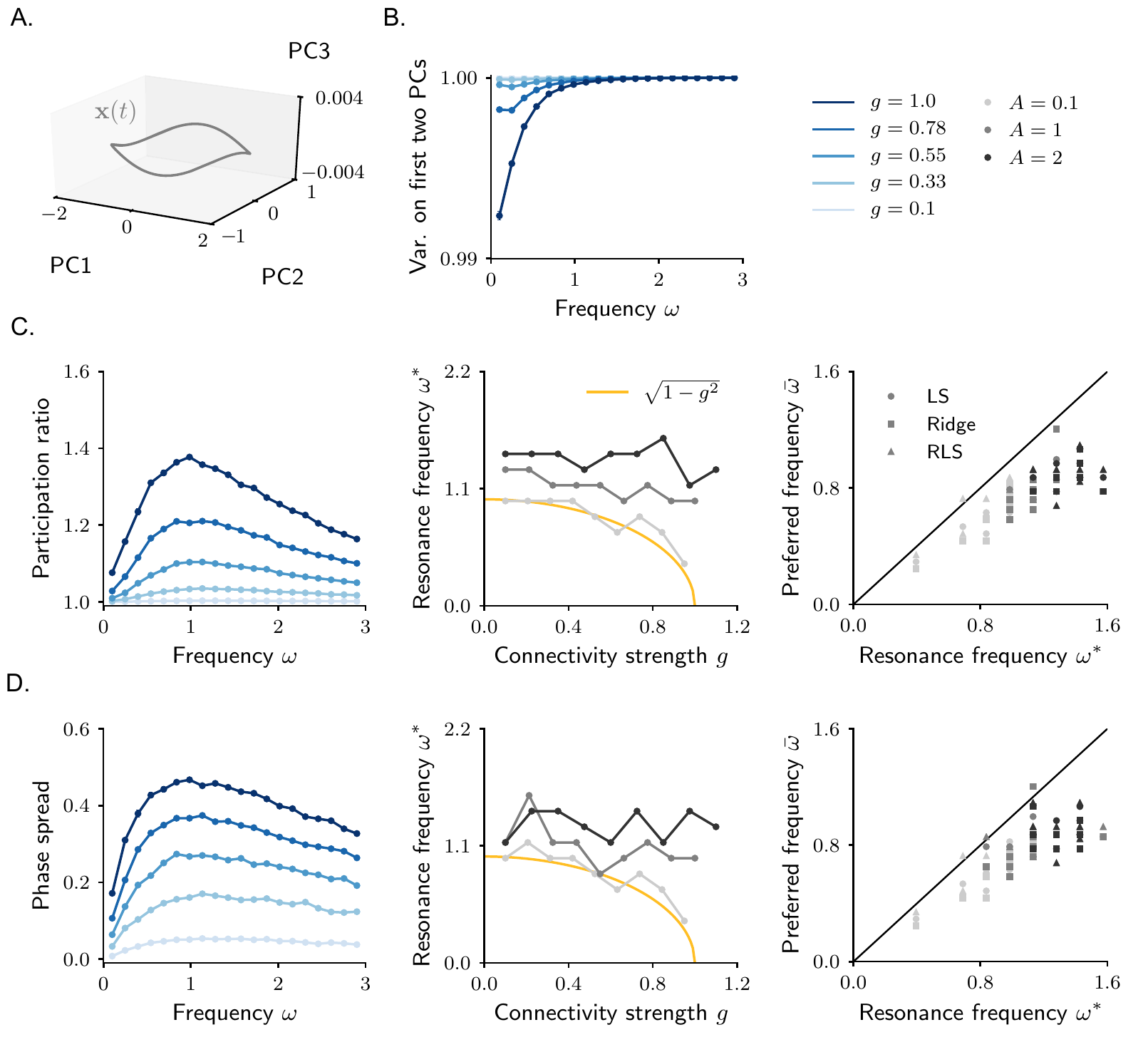}
		\caption{
			\textbf{Using open-loop reservoir dynamics to predict training performance in non-linear networks.}
			{\bf A.} Reservoir trajectories $\mathbf{x}(t)$ in driven non-linear networks:  example trajectory projected onto the first three PCs of network activity (note scale of PC3 axis).  
			{\bf B.} Variance explained by projecting trajectories on the first two PC axes (note scale of variance). 
			{\bf C.} Using representation dimensionality to predict training performance.
			Left: participation ratio of driven trajectory as a function of $\omega$ for a range of $g$ values (blue shades). 
			Results are averages over 20 simulations of networks of size $N=2000$, with $A=1$.
			Center: resonance frequency $\omega^*$, measured as the position of maximum dimensionality, as a function of $g$ for three different values of $A$ (grey shades).
			Right: error-minimizing frequency $\bar{\omega}$ (from Fig.~\ref{fig:fig1}) plotted against $\omega^*$ (from center panel). Results are shown for three training algorithms (legend). 
			{\bf D.} Using phase spread of driven trajectories to predict performance. 
			Unit activities $x_i(t)$ were fitted with sinusoidal functions of the driving frequency $\omega$, and the variance of the phase distribution was measured. Left, center and right panels are the same as in \textbf{C}. 
			\label{fig:fig6}
		}
	\end{figure}

	Guided by these insights, we examined the properties of open-loop dynamics (Eq.~\eqref{eq:dyn_reservoir}) in non-linear networks. 
	Because of the non-linearity, the neural trajectory $\mathbf{x}(t)$ is in this case not planar, but curved along many dimensions (Fig.~\ref{fig:fig6}{\bf A}); most of its variance, however, is still explained by two directions (Fig.~\ref{fig:fig6}{\bf B}).
	We investigated numerically the properties of non-linear target representations by using the same measures as for linear networks, namely the dimensionality and the spread of phases across units.
	Although in non-linear systems these are not equivalent measures, we find that their behaviour is qualitatively similar to one another, and to the behaviour of their analogues in linear systems (Fig.~\ref{fig:fig6}{\bf C-D} left). 
	First, both measures increase monotonically with the connectivity strength $g$. 
	Second, for any fixed value of $g$, both measures display a maximum at an intermediate frequency $\omega^*$.
	
	In the middle panels of Fig.~\ref{fig:fig6}{\bf C-D}, we display the resonance frequency $\omega^*$ computed from both measures of non-linear representations (left panels) across various values of $g$ and for three target amplitudes (legend).
	As in the linear case, we find that the value of $\omega^*$ decreases with the connectivity strength $g$; unlike the linear case, however, it 
	depends on the target amplitude $A$.
	For small target amplitudes (light gray), 
	both measures of non-linear representations yield values of $\omega^*$ which are quantitatively very close to the values predicted by the linear theory, i.e.~$\sqrt{1-g^2}$ (yellow line). 
	This is expected, as for low-amplitude driving the reservoir activity mostly remains in the vicinity of the origin, a region where the non-linear dynamics are approximately linear. 
	In the nonlinear case, however, as the target amplitude $A$ increases (darker shades of gray, see legend), the resonance frequency $\omega^*$ also increases. The decrease of $\omega^*$ with $g$ is retained, but to a lesser extent. 
	
	In the right panels of Fig.~\ref{fig:fig6}{\bf C-D}, we compare the resonance frequency $\omega^*$ predicted from analyzing non-linear representations to the preferred frequency $\bar{\omega}$ which minimizes training performance  (Fig.~\ref{fig:fig1}).
	Although the two quantities do not exactly coincide, they display significant correlations.
	Remarkably, the value of $\omega^*$ correctly captures the behaviour of the preferred frequency $\bar{\omega}$ with the target amplitude $A$: like $\omega^*$, $\bar{\omega}$ increases with $A$, as can be seen by the clustering of different shades of grey in Fig.~\ref{fig:fig6}{\bf C-D}, right panels. 
	
\begin{figure}
		
		\centering
		\includegraphics{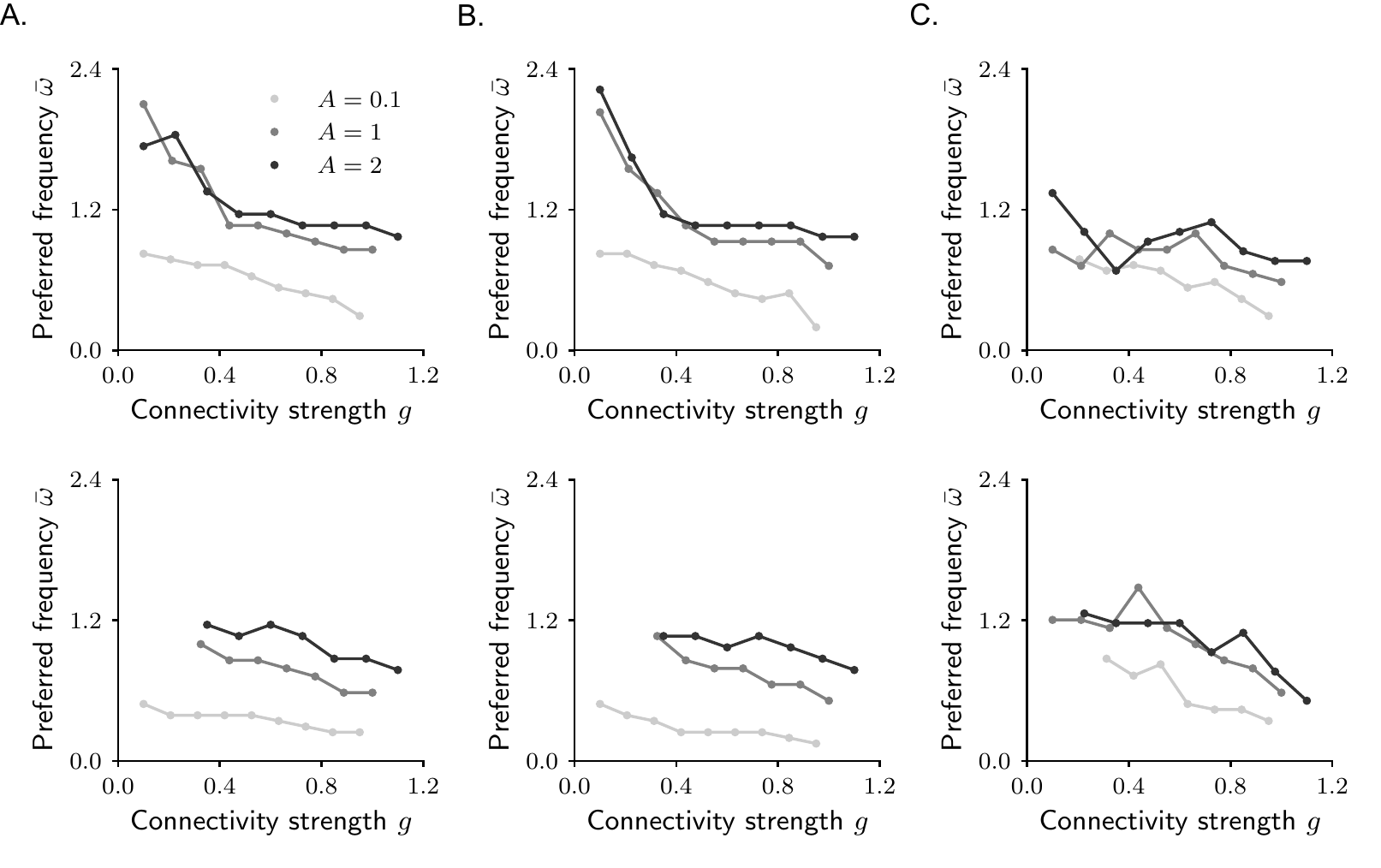}
		\caption{
			\textbf{Emergence of preferred frequency in non-linear feedback networks, supplementary results obtained for different hyper-parameters.}
			{\bf A.} LS training. Parameters are as in Fig.~\ref{fig:fig1}, except $\sigma^{\text{pert}}=0.01$ (top) and $\sigma^{\text{LS}}=0.001$ (bottom).
			{\bf B.} Ridge training. Parameters are as in Fig.~\ref{fig:fig1}, except $\sigma^{\text{pert}}=0.01$ (top) and  $\sigma^{\text{R}}=0.1$ (bottom).
			{\bf C.} RLS training. Parameters are as in Fig.~\ref{fig:fig1}, except $N^{\text{tot}}=10$ (top and bottom), $\sigma^{\text{pert}}=0.01$ (top) and $\alpha=5$ (bottom). In the bottom row, we have removed from the plot the preferred frequencies $\bar{\omega}$ at low $g$ values in the cases where training fails (i.e.~it is characterized by very high error) for every value of $\omega$ tested.
			\label{fig:S1}
		}
	\end{figure}

	Importantly, this observation is not sensitive to training details.
	It is consistent across the three training algorithms we used (Fig.~\ref{fig:fig6}{\bf C-D}, right panels) and across a broad range of training hyper-parameters (Supp.~Fig.~\ref{fig:S1}). 
	These results suggest that the properties of the internal representation -- here measured by the dimensionality and synchronization of the open-loop dynamics -- play a crucial role in determining training performance of non-linear networks, as they do for linear networks.
	
	\section{Discussion}
	
	Ubiquitously across biology, complex high-dimensional systems interact with their environment through low-dimensional channels.
	The computational modelling of such setups has advanced considerably in the past two decades with the emergence of reservoir computing techniques \cite{JaegerHaas, Maass2002}, where learning acts on such low-dimensional bottlenecks.
	Despite the simplicity of this learning scheme, the factors contributing to or hindering the success of training in reservoir networks are in general not well understood \cite{JaegerRiddle}.
	In particular, a theory is lacking for predicting -- based on the characteristics of the reservoir and the target function -- dynamics and performance of trained feedback networks.
	
	In this work, we studied learning performance of feedback networks trained to self-sustain a sinusoidal readout signal. 
	Through mathematical analysis, we showed that learning performance is mostly controlled by the quality of the internal representation of the target signal. 
	This quality can be quantified by analyzing the open-loop dynamics and measuring the condition number of their cross-correlation matrix, a number that characterizes to what extent the network dynamics is robust to training noise.
	We found that the condition number displays, like training performance, a complex dependence on the parameters controlling the reservoir internal properties (strength of reservoir connectivity $g$) and the readout target function (frequency $\omega$). 
	The parameter values where the condition number is minimized,  $\omega^*=\sqrt{1-g^2}$, define an optimal spot for learning. 
	At this optimal point, internal representations are characterized by maximal dimensionality and minimal synchrony, which are two ways of quantifying the richness of the dynamic repertoire available to the learning algorithm. 
	Our insights were derived by studying linearized dynamics and were later tested on non-linear networks, where they successfully capture non-trivial aspects of training performance.
	
	The condition number of the cross-correlation matrix has been pointed out in several studies as a key quantity in determining performance \cite{JaegerRiddle, Lukosevicius}. 
	Our work analytically quantifies those empirical observations in the framework of networks trained on a simple task via common LS-based algorithms. 
	We have shown, however, that performance might depend on other features, such as closed-loop stability, for other non-standard algorithms (see Fig.~\ref{fig:ClosedLoop}). 
	
	Importantly, our analysis differentiates between two properties that might hinder network performance: high-norm readouts and non-normality.
	In a number of classic studies \cite{SussilloAbbott,JaegerRiddle,Schuessler2020}, large norms of the readout vector have been associated with impaired performance.
	In addition, recent observations indicate that training performance is low in parameter regions where the open-loop dynamics is highly non-normal \cite{Logiaco2019}, and link low performance to large readout vectors. 
	In our framework, the two properties can be analyzed separately.
	Non-normality can be measured from the angle $\theta$ between the two activity eigenvectors $\mathbf{v}_\pm$; Fig.~\ref{fig:Encoding_Trajectories}{\bf C} indicates that non-normality is minimal at the resonance frequency $\omega^*$. The norm of the readout vector $\mathbf{n}_{LS}$ can be instead derived from Eq.~\eqref{eq:readout-eq} (see Appendix \ref{app:LSnorm}); we show in Supp.~Fig.~\ref{fig:Norm_n} that, for every value of connectivity $g$, the norm of the readout vector is monotonic in the target frequency $\omega$.
	We conclude that these two quantities are not equivalent predictors of learning performance; in our setting, training performance is optimal close to $\omega^*$, so that non-normality is identified as the dominating factor in controlling performance.
	
	Several studies have supported the hypothesis that learning capability is maximized in the parameter region where dynamics is close to the boundary between ordered and chaotic activity, i.e.~the \emph{edge-of-chaos} \cite{LegensteinMaass, SussilloAbbott, LukoseviciusJaeger}. 
	Our findings are consistent with this hypothesis: we have shown that the condition number (and, consequently, the training error) monotonically decreases as the strength of reservoir connectivity $g$ is increased from 0 towards its critical value. 
	However, our analysis has shown that, together with the strength of internal connectivity, learning performance is crucially shaped by the properties of the target function.
	By analysing non-linear networks, furthermore, we have found that the parameter region characterized by maximally high-dimensional and de-synchronized internal representations does not necessarily coincide with the edge-of-chaos; the two regions in fact diverge as the target amplitude $A$ is increased and activity becomes strongly non-linear (Figs.~\ref{fig:fig6} and \ref{fig:EOC}). 
	Specifically, as $A$ increases, the critical frequency where activity becomes chaotic moves to very high values \cite{Rajan2010} (Fig.~\ref{fig:EOC}), while the resonance frequency $\omega^*$ (which measures activity dimensionality and synchrony) remains close to the training-preferred frequency $\bar{\omega}$ (Fig.~\ref{fig:fig6}). 
	This result suggests that future research should focus on characterizing the properties of driven non-linear activity rather than analysing the transition to chaos per-se.
	
	
	The numerical analysis of non-linear networks (Fig.~\ref{fig:fig6}), which was led by the insights gained from the linear theory, suggests that representation quality is a major determinant of closed-loop performance also in the case of non-linear networks.
	Exploiting the link between the two, we were able to predict the dependence of the preferred frequency $\bar{\omega}$ on both the internal connectivity $g$ and the target amplitude $A$, which plays no role in the linear counterpart.
	This is despite the fact that the non-linearity of the dynamics introduces, in trained networks, new qualitative behaviours which do not exist in linear networks.
	In particular, we observe that the training error (and, consequently, the value of $\bar{\omega}$) strongly depends on the hyper-parameters controlling the stability of the limit cycle which constitutes the internal representation (see Appendix \ref{app:training} and Supp.~Fig.~\ref{fig:S2}).
	In this respect, a more detailed analysis is called for; we hope that future work would extend our analytic framework to cover non-linear reservoirs.
	
	\bibliographystyle{unsrt} 
	\bibliography{main} 
	
	\subsection*{Acknowledgments}
	This work was supported in part by the Israeli Science Foundation (grant number 346/16, OB; and grant number 155/18, NB). FM would like to thank Haim Sompolinsky and Ran Rubin for their supervision on an early version of the project. FM would also like to thank Laureline Logiaco for useful discussions, and Srdjan Ostojic and Manuel Beiran for their feedback on a previous version of the manuscript.
	LS would like to thank Friedrich Schuessler for helpful discussions.

	\newpage
	
	\section{Appendix}

	\subsection{Training of feedback networks}
	\label{app:training}
	
	In the following, we report the procedures used to train feedback architectures (Figs.~\ref{fig:fig1} and \ref{fig:predicting_perf}). Procedures are detailed for the general case of non-linear networks; the case of linear networks corresponds to taking $\Phi(x)=x$. 
	Results are averages across 1000 different network and training realizations.
	
	\paragraph{LS regression training} Training is performed in the open-loop setup. In a first phase, open-loop activity (where we enforce $u(t)=A\cos(\omega t)$ in Eq.~\eqref{eq:dyn_reservoir}) is simulated by using the Scipy \textsf{odeint} routine from $t=0$ to $t=T^{\text{tot}}$, with $T= N^{\text{tot}} 2 \pi / \omega$. Activity is stored in a $L \times N$ matrix $\bm{\Phi}$, where $L$ indicates the number of time points used for integration. The $L$-dimensional vector $\bm{F}$ is constructed by computing the target function $f(t)$ at the same time points.
	Activity and target function from $t=0$ to $T^{\text{tr}} = N^{\text{tr}} 2 \pi / \omega$ are later discarded, resulting in $L' \times N$ and $L' \times 1$ matrices $\bm{\Phi}$ and $\bm{F}$, where $L'$ indicates the number of time points kept after discarding the transient.
	We used $N^{\text{tot}}=20$ and $N^{\text{tr}}=8$. 
	In order to regularize the cross-correlation matrix and to ease local stability in non-linear networks, white noise is then added on top of activity: $\bm{\tilde{\Phi}} = \bm{\Phi} + \sigma^{\text{LS}} \bm{\xi}$,
	where $\bm{\xi}$ is a $L' \times N$ matrix of standard Gaussian variables.
	The trained readout vector $\mathbf{n}$ is finally computed as:
	\begin{equation}
	\mathbf{n} = (\bm{\tilde{\Phi}}^\top\bm{\tilde{\Phi}})^{-1} \bm{\tilde{\Phi}}^\top \bm{F}.
	\end{equation}
	
	In linear networks, training performance is measured in terms of the mismatch between the target outlier eigenvalues $\lambda_\pm$ (see Section \ref{sec:closed_loop}) and the outlier eigenvalues $\tilde{\lambda}_\pm$, defined as the pair of complex conjugate eigenvalues of $\bar{\mathbf{J}}=\mathbf{J}+\mathbf{m}\mathbf{n}^\top$ whose real part is maximally close to one.
	In non-linear networks, performance is measured on closed-loop activity. 
	To this end, closed loop dynamics (Eq.~\eqref{eq:dyn_feedback}) is simulated from $t=0$ to $t=T^{\text{tot}}$. 
	The initial condition is taken to be equal to activity in the last time step of the open-loop simulation; on top of it, an $N$-dimensional vector of white noise of amplitude $\sigma^{\text{pert}} A$ is added. 
	The latter perturbation was used to take into account training error generated by unstable local dynamics; we take $\sigma^{\text{pert}}=0$ in linear networks.
	To measure the test error we fitted a sinusoidal function $F(t)$ of fixed amplitude $A$ and frequency $\omega$ to the readout signal $\mathbf{z}=\mathbf{n}^\top \bm{\Phi}^\top$ obtained in the closed-loop simulation, yielding a novel $L$-dimensional vector $\bm{F}$. Readout error is finally measured as $\langle \: | z_k - F_k |\: \rangle_k$, where the average is taken over all the integration time points from $t=0$ to $t=T^{\text{tot}}$. 
	If the fit fails, we set the readout error to 1.
	Parameters used in Fig.~\ref{fig:fig1} are $N=400$, $\sigma^{\text{LS}}=0.01$ and $\sigma^{\text{pert}}=0.1$.
	Parameters used in Fig.~\ref{fig:predicting_perf} are $N=400$, $\sigma^{\text{LS}}=0.01$ and $\sigma^{\text{pert}}=0$.
	
	\paragraph{Ridge regression training} As in the LS case, training is performed in the open-loop setup. The $L'\times N$ open-loop activity $\bm{\Phi}$ matrix is obtained as above.
	The trained readout vector $\mathbf{n}$ is then computed as:
	\begin{equation}
	\mathbf{n} = (\bm{{\Phi}}^\top\bm{{\Phi}} + (\sigma^{\text{R}})^2 \mathbf{I})^{-1} \bm{{\Phi}}^\top \bm{F}
	\end{equation}
	where $\mathbf{I}$ indicates the $N$-dimensional identity matrix. Training performance is measured as in the LS case.
	Parameters used in Fig.~\ref{fig:fig1} are $N=400$, $\sigma^{\text{R}}=1$ and $\sigma^{\text{pert}}=0.1$.
	Parameters used in Fig.~\ref{fig:predicting_perf} are $N=400$, $\left(\sigma^{\text{R}}\right)^2=L'/2 \cdot \sigma^2$ and $\sigma^{\text{pert}}=0$; $\sigma^2 = 10^{-7}$ is the regularization parameter used for regression in the Fourier space (see Appendix \ref{app:ridge_regression}), which was used to compute the theoretical prediction for outlier eigenvalues.
	
	\paragraph{RLS training} Training is performed in the closed-loop setup, from $t=0$ to $t=T^{\text{tot}}$, with $T= N^{\text{tot}} 2 \pi / \omega$ and $N^{\text{tot}}=20$ (Fig.~\ref{fig:fig1}) or 10 (Fig.~\ref{fig:predicting_perf}). 
	At $t=0$, an $N\times N$-dimensional matrix $\mathbf{P}$ is initialized as: $\bm{P}=\mathbf{I}/\alpha$, where $\mathbf{I}$ indicates the $N$-dimensional identity matrix and $\alpha$ is a free parameter. Matrix $\mathbf{P}$ represents a running estimate of the inverse of the activity cross-correlation matrix \cite{SussilloAbbott}. 
	Readout vector $\mathbf{n}$ is further initialized with zero entries.
	At every learning step, closed-loop activity is simulated from $t_0$ to $t_0+\tau$ (Eq.~\eqref{eq:dyn_feedback}), with $\tau = (2\pi / \omega)/500$. Activity at $t=t_0+\tau$ is stored in an $N$-dimensional vector $\bm{\Phi}$. Matrix $\mathbf{P}$ is then updated as \cite{SussilloAbbott}:
	\begin{equation}
	\mathbf{P} \leftarrow \mathbf{P} - \frac{ \mathbf{P} \bm{\Phi}^\top \bm{\Phi}\mathbf{P} }{1 + \bm{\Phi}^\top \mathbf{P} \bm{\Phi}}
	\end{equation}
	The readout vector $\mathbf{n}$ is then updated as:
	\begin{equation}
	\mathbf{n} \leftarrow \mathbf{n} - e \mathbf{P} \bm{\Phi}
	\end{equation}
	where the error $e$ is measured as: $e=z(t_0+\tau)-f(t_0+\tau)$.
	Once training is completed, performance is measured as in the LS case. 
	Parameters used in Fig.~\ref{fig:fig1} are $N=400$, $\alpha=1$ and $\sigma^{\text{pert}}=0.1$.
	Parameters used in Fig.~\ref{fig:predicting_perf} are $N=400$, $\alpha=1$ and $\sigma^{\text{pert}}=0$.
	
	\begin{figure}
		
		\centering
		\includegraphics{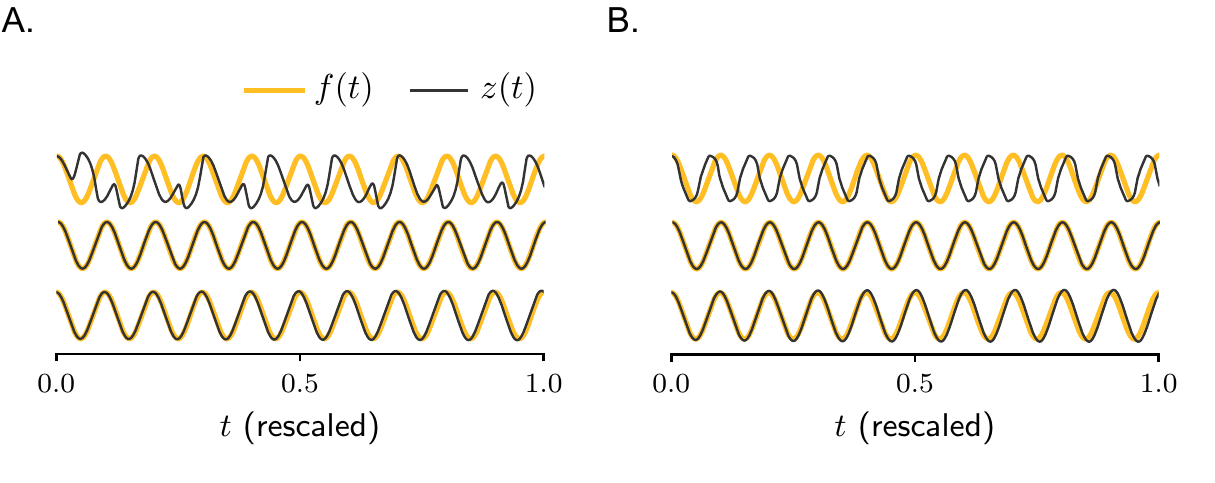}
		\caption{
			\textbf{Emergence of preferred frequency in non-linear feedback networks, example trials.}
			Example trials as in Fig.~\ref{fig:fig1}{\bf B} where training is performed via Ridge regression ({\bf A}) or RLS ({\bf B}). Parameters are as in Fig.~\ref{fig:fig1}{\bf B}, except  $N^{\text{tot}}=3$ in {\bf B}.
			\label{fig:S2}
		}
	\end{figure}
	
	\subsection{Analysis of linear open-loop reservoirs}
	\label{app:linear_ODE}
	
	For a general input $f(t)$, the system of linear equations Eq.~\eqref{eq:linear_ODE} admits the asymptotic solution ($t\rightarrow\infty$)
	\begin{equation}
	\mathbf{x}(t) = \int_0^t e^{\mathbf{(J-I)}(t-\tau)} \: \mathbf{m}f(\tau)\: \diff\tau,
	\end{equation}
	which for a complex exponential $f(t) = e^{st}$ for $s \in \mathbb{C}$, simplifies to
	\begin{equation}
	\mathbf{x}(t) = [(1+s)\mathbf{I} - \mathbf{J}]^{-1}\mathbf{m}e^{st}.
	\end{equation}
	In particular, for $f(t)=\cos(\omega t) =\frac{1}{2}(e^{i\omega t}+e^{-i\omega t})$, one finds the expression in Eq.~\eqref{eq:x+-} in the main text. 
	Note that $\mathbf{x}_{\pm}$ defined in Eq.~\eqref{eq:x+-} 
	correspond to the amplitude of the peaks of the Fourier transform of $\mathbf{x}(t)$; indeed, we have:
	\begin{equation}
	\begin{split}
	\mathbf{\hat{x}} (\hat{\omega}) &= \int_{-\infty}^{\infty} \mathbf{x}(t) e^{-i\hat{\omega}t} \diff t = \frac{1}{2} [\mathbf{x_+} \delta(\hat{\omega}- \omega) + \mathbf{x_-} \delta(\hat{\omega} + \omega)].
	\end{split}
	\end{equation}
	
	In deriving Eq.~\eqref{eq:driven_state}, we defined:
	\begin{equation}\label{eq:xtov}
	\begin{split}
	\mathbf{v}_+ &:= \frac{1}{2}\left(\mathbf{x}_{+} + \mathbf{x}_{-} \right) \\
	\mathbf{v}_- &:= \frac{i}{2}\left(\mathbf{x}_{+} - \mathbf{x}_{-} \right).
	\end{split}
	\end{equation}

	\subsection{Statistics of spanning vectors $\mathbf{v}_+$ and $\mathbf{v}_-$}
	\label{app:spanning_vecs}
	
	In this section, we characterize the geometry of vectors $\mathbf{v}_+$ and $\mathbf{v}_-$ in terms of their norms and overlap.
	
	We start by evaluating the dot product:
	\begin{equation}\label{eq:dot_prod}
	\left[\left(\mathbf{I} - \frac{\mathbf{J}}{a}\right)^{-1}\mathbf{m}\right] \cdot \left[\left(\mathbf{I} - \frac{\mathbf{J}}{b}\right)^{-1}\mathbf{m}\right] =
	\mathbf{m}^\top \left[\left(\mathbf{I} - \frac{\mathbf{J}}{a}\right)^{-1}\right]^\top \left(\mathbf{I} - \frac{\mathbf{J}}{b}\right)^{-1}\mathbf{m}
	\end{equation}
	with $a$, $b\in \mathbb{C}$. 
	If the eigenvalues of $\mathbf{J}/a$ and $\mathbf{J}/b$ have real part smaller than one, we can use the power series expansion
	\begin{equation}
	\left(\mathbf{I} - \frac{\mathbf{J}}{a}\right)^{-1} = \sum_{p=0}^{\infty} \frac{\mathbf{J}^p}{a^p},
	\end{equation}
	so that Eq.~\eqref{eq:dot_prod} becomes
	\begin{equation}
	\mathbf{m}^T \sum_{p=0}^{\infty}\sum_{q=0}^{\infty}\frac{\left(\mathbf{J}^p\right)^\top \mathbf{J}^q}{a^p b^q} \mathbf{m}.
	\end{equation}
	
	Since $\mathbf{J}$ is random, the value of this expression randomly fluctuates across different realizations of recurrent connectivity $\mathbf{J}$. We thus turn to a statistical characterization, and evaluate Eq.~\eqref{eq:dot_prod} by computing its mean and variance with respect to different realizations of $\mathbf{J}$.
	
	The mean yields, to the leading order in $N$ \cite{Schuessler2020}:
	\begin{equation}\label{eq:dot_prod_mean}
	\begin{split}
	&\mathbf{m}^T \sum_{p=0}^{\infty}\sum_{q=0}^{\infty}\frac{\left\langle(\mathbf{J}^p)^\top\mathbf{J}^q\right\rangle_{\mathbf{J}}}{a^p b^q} \mathbf{m} \\
	&= \mathbf{m}^T\mathbf{m}  \sum_{p=0}^{\infty}\left(\frac{g^2}{ab}\right)^p  \\
	&= N \sum_{p=0}^{\infty}\left(\frac{g^2}{ab}\right)^p\\
	&= N\frac{1}{1-\frac{g^2}{ab}} = N \frac{ab}{ab - g^2}.
	\end{split}
	\end{equation}
	We have used:
	\begin{equation}\label{eq:random_m}
	\left\langle(\mathbf{J}^p)^T\mathbf{J}^q\right\rangle_{\mathbf{J}}=\delta_{pq} \: \mathbf{I} \: {g^{2p}}
	\end{equation}
	which comes from observing that $\mathbf{J}^p$ is a random matrix, which is uncorrelated to $\mathbf{J}^q$ for $q\neq p$ has variance $g^{2p}/N$. 
	This yields:
	\begin{equation}
	\sum_{k=1}^N \langle (J^p)_{ki}(J^q)_{kj}\rangle_{\mathbf{J}} = \sum_{k=1}^N \delta_{ij} \delta_{pq} \frac{g^{2p}}{N} =  \delta_{ij} \delta_{pq} g^{2p},
	\end{equation}
	from which we obtain Eq.~\eqref{eq:random_m}.
	
	The variance can be computed in a similar way. Like the mean, the variance is characterized by $\mathcal{O}(N)$ scaling \cite{Schuessler2020}; as a consequence, variability due to different realizations of $\mathbf{J}$ does not enter the dot product Eq.~\eqref{eq:dot_prod} to the leading order in $N$, and dot products can be replaced with their mean (Eq.~\eqref{eq:dot_prod_mean}) when $N\rightarrow\infty$.
	
	We can now compute the mean norm of the spanning vectors from combining Eqs.~\eqref{eq:xtov} and \eqref{eq:dot_prod_mean}:
	\begin{equation}
	\begin{split}
	\frac{\|\mathbf{v}_+ \|^2}{N} &= \frac{1}{4N} \left(\mathbf{x}_+ \cdot \mathbf{x}_+ + 2 \mathbf{x}_+ \cdot \mathbf{x}_-+ \mathbf{x}_-\cdot \mathbf{x}_-\right)\\
	&= \frac{1}{4}\left(\frac{1}{(1+i\omega)^2 - g^2} + \frac{2}{1+\omega^2 - g^2} + \frac{1}{(1-i\omega)^2 - g^2} \right)\\
	&= \frac{(1 - g^2)^2 + \omega^2(2-(1-g^2))}{((1-g^2 -\omega^2)^2 + 4\omega^2)(1-g^2+\omega^2)};
	\end{split}
	\end{equation}
	the calculation of the norm of $\mathbf{v}_-$ is very similar, and only differs in the sign of the first summand in the second line above:
	\begin{equation}
	\begin{split}
	\frac{\|\mathbf{v}_- \|^2}{N} &= \frac{1}{4}\left(-\frac{2(1 - g^2 -\omega^2)}{(1-g^2 -\omega^2)^2 + 4\omega^2} + \frac{2}{1 - g^2+\omega^2} \right)\\
	&= \frac{\omega^2 (\omega^2 + 2-(1-g^2))}{((1-g^2 -\omega^2)^2 + 4\omega^2)(1-g^2+\omega^2)}.
	\end{split}
	\end{equation}
	Finally, when computing the dot product between the two vectors, the cross terms cancel to yield
	\begin{equation}
	\begin{split}
	\frac{\mathbf{v}_+ \cdot \mathbf{v}_-}{N} &= \frac{1}{4}\left(-\frac{1}{(1+i\omega)^2 - g^2} + \frac{1}{(1-i\omega)^2 - g^2} \right)\\
	&= \frac{\omega}{(1-g^2 -\omega^2)^2 + 4\omega^2}.
	\end{split}
	\end{equation}
	With these expressions, the angle $\theta$ between $\mathbf{v}_+$ and $\mathbf{v}_-$ can be written as
	\begin{equation}
	\text{cos}(\theta) = \frac{1-g^2+\omega^2}{\sqrt{(1 - g^2)^2 + \omega^2(2-(1-g^2))}\sqrt{\omega^2 + 2-(1-g^2)}},
	\end{equation}
	
	By denoting $\varepsilon = 1-g^2$, we now summarize the statistics of the spanning vectors $\mathbf{v}_\pm$:
	\begin{equation} \label{eq:geometric_qs}
	\boxed{\begin{split}
		&\frac{\|\mathbf{v}_+ \|^2}{N} = \frac{\omega^2(2-\varepsilon) + \varepsilon^2}{(\varepsilon+\omega^2)((\varepsilon-\omega^2)^2 + 4\omega^2)} \\
		&\frac{\|\mathbf{v}_- \|^2}{N} = \frac{\omega^2(2-\varepsilon+\omega^2)}{(\varepsilon+\omega^2)((\varepsilon-\omega^2)^2 + 4\omega^2)}\\
		&\frac{\mathbf{v}_+ \cdot \mathbf{v}_-}{N} = \frac{\omega}{(\varepsilon-\omega^2)^2 + 4\omega^2}
		\end{split}}
	\end{equation}
	and angle between the two spanning vectors reads:
	\begin{equation} \label{eq:costheta_explicit}
	\boxed{\text{cos}(\theta) = \frac{\mathbf{v}_+ \cdot \mathbf{v}_-}{\|\mathbf{v}_+ \|\|\mathbf{v}_- \|} = \frac{\varepsilon+\omega^2}{\sqrt{2-\varepsilon+\omega^2}\sqrt{\varepsilon^2 + \omega^2(2-\varepsilon)}}.}
	\end{equation}
	
	\subsection{Analysis of geometric properties of driven trajectories}
	\label{app:geometric_analysis}
	
	We can use the expressions computed in Appendix \ref{app:spanning_vecs} to evaluate the participation ratio $d$:
	\begin{equation}
	d = \frac{(\nu_1+\nu_2)^2}{\nu_1^2+\nu_2^2}
	\end{equation}
	where $\nu_1$ and $\nu_2$ are the eigenvalues of the reduced cross-correlation matrix $C^R$ (Eq.~\eqref{eq:Amatrix}). 
	We express the eigenvalues as
	\begin{equation}
	\nu_{1,2} = \frac{1}{2}\left(  \gamma \pm \sqrt{\Delta^2} \right),
	\end{equation}
	where $\gamma := \text{Tr} (C^R)$ and $\Delta^2 := \text{Tr}^2(C^R) - 4 \cdot \text{det} (C^R)$. This yields:
	\begin{equation}
	d = 2 \frac{\gamma^2}{\gamma^2 + \Delta^2}
	\end{equation}
	from which Eq.~\eqref{eq:participation_ratio} follows.
	Inserting the results from Appendix \ref{app:spanning_vecs}, we also have:
	\begin{equation} \label{eq:PR_explicit}
	\boxed{d = \frac{\varepsilon^2 - 2\varepsilon\omega^2 + 4\omega^2 + \omega^4}{\varepsilon^2 + 2\omega^2 + \omega^4}.}
	\end{equation}
	
	We now consider some limiting behaviours of the quantities computed above.
	In the high-frequency limit $\omega\to \infty$, the norms and dot product of $\mathbf{v}_{\pm}$ vanish, whereas the overlap is given by
	\begin{equation}
	\lim_{\omega\to\infty}\text{cos}(\theta) = \frac{1}{\sqrt{2-\varepsilon}} = \frac{1}{\sqrt{1+g^2}};
	\end{equation}
	the dimensionality $d$ tends to 1 in this limit.
	
	In the limit of low frequency, for any $0<g<1$ the norms and dot product can be evaluated directly by substituting $\omega = 0$ to obtain
	\begin{equation}
	\|\mathbf{v}_- \| = \mathbf{v}_+ \cdot \mathbf{v}_- = 0
	\end{equation}
	and
	\begin{equation}
	\|\mathbf{v}_+ \| = \frac{1}{\sqrt{2\left(1-g^2\right)}},
	\end{equation}
	while the overlap assumes the same non-zero value as in the other extreme: 
	\begin{equation}\label{eq:costheta_w0}
	\text{cos}(\theta) = \frac{1}{\sqrt{1+g^2}}
	\end{equation}
	and the dimensionality is one.
	
	In between the two extremes, the dot product, the participation ratio, and the norm $\|\mathbf{v}_- \|$ obtain a maximum value at a finite frequency.
	For any $g$, we compute the frequency yielding the minimum angle, by searching for local extrema of Eq.~\eqref{eq:costheta_explicit}:
	\begin{equation}
	\frac{\partial \text{cos}^2(\theta)}{\partial \omega} = \frac{\partial}{\partial\omega}\frac{\left(\omega^2+\varepsilon\right)^2}{\left(2-\varepsilon\right) \left(\omega^4 - 2\omega^2\varepsilon + \varepsilon^2\right) + 4\omega^2} = 0,
	\end{equation}
	the numerator of which, after some algebra, can be written as $8\omega\left(\omega^4 - \varepsilon^2\right)$.
	The angle thus has a local minimum at
	\begin{equation}
	\omega^* = \sqrt{\varepsilon} = \sqrt{1-g^2}.
	\end{equation}
	Likewise, from Eq.~\eqref{eq:PR_explicit} we have
	\begin{equation}
	\frac{\partial d}{\partial \omega} = 4\frac{\omega(\varepsilon - 1)\left[ \omega^4 - \varepsilon^2  \right]}{(\varepsilon^2 + 2\omega^2 + \omega^4)^2},
	\end{equation}
	which, again, implies a nontrivial maximum of $d$ at $\omega^*(g) = \sqrt{1-g^2}$.
	
	Finally, we observe that $g=1$ is a singular point, as in the limit of low frequency we have
	\begin{equation}
	\lim_{\omega\to 0} \|\mathbf{v}_+ \| = \lim_{\omega\to 0} \|\mathbf{v}_- \|= \lim_{\omega\to 0} \mathbf{v}_+ \cdot \mathbf{v}_- = \infty.
	\end{equation}
	but the overlap vanishes
	\begin{equation}\label{eq:costheta_g1}
	\lim_{\omega\to 0}\text{cos}(\theta) = 0,
	\end{equation}
	and the participation ratio attains its global maximum
	\begin{equation}
	\lim_{\omega\to 0} d = 2.
	\end{equation}
	Note that the limits $\lim_{g\to 1, \omega\to 0} d$ and $\lim_{g\to 1, \omega\to 0}\text{cos}(\theta)$ do not exist, since they depend on the order of limits taken.
	To see this, compare Eqs. \eqref{eq:costheta_w0} and \eqref{eq:costheta_g1}.

	\subsection{Distribution of response phases}
	\label{app:phases}
	The phase spread in response of different units (Eq.~\eqref{eq:phasespread}) was computed as a numerical integral performed over the probability distribution $p(\phi)$ whose analytical form is available in \cite{Aalo2007}.
	We used:
	\begin{equation}
	p(\phi) = \frac{1}{\pi \| \mathbf{v}_+ \| \| \mathbf{v}_-  \| \sqrt{1-\rho^2}}
	\left[
	\frac{\cos^2(\phi)}{\| \mathbf{v}_+ \|^2 (1-\rho^2)} + 
	\frac{\sin^2(\phi)}{\| \mathbf{v}_- \|^2 (1-\rho^2)} 
	- \frac{ 2 \rho \sin(\phi) \cos(\phi)}{\| \mathbf{v}_+ \|\| \mathbf{v}_- \| (1-\rho^2)} 
	\right]^{-1}
	\end{equation}
	where $\rho=\cos(\theta)$.
	From \cite{Aalo2007} we also used:
	\begin{equation}
	\bar{\phi} = \frac{1}{2} \arctan \left( \frac{2 \rho \| \mathbf{v}_+ \|\| \mathbf{v}_- \|}{\| \mathbf{v}_+ \|^2 - \| \mathbf{v}_- \|^2} \right).
	\end{equation}
	
	\subsection{\emph{From-k} regression}
	\label{app:from-k}
	
	In this section, we explain how \emph{from-k} least-squares regression (Fig.~\ref{fig:ClosedLoop}) is performed.
	
	For $2\le k \le N$, we define the cropped spanning vectors as
	\begin{equation}
	\left [\mathbf{v}_\pm^k \right]_i = \begin{cases}
	\left[\mathbf{v}_\pm \right]_i, & \text{if}\ 1 \le i\le k \\
	0, & \text{otherwise}
	\end{cases}
	\end{equation}
	where $\left[\mathbf{v}_\pm \right]_i$ indicates the $i$-th element of vectors $\mathbf{v}_\pm$.
	For every $k$, the $\emph{from-k}$ LS regressor is given by the pseudo-inverse, which yields the LS regressor of minimum norm:
	\begin{equation}
	\begin{aligned}
	\mathbf{n}^k_{LS} = 
	\begin{pmatrix}
	\mathbf{v}_+^k & \mathbf{v}_-^k
	\end{pmatrix}
	\begin{pmatrix}
	\| \mathbf{v}_+^k \|^2 & \mathbf{v}_+^k\cdot \mathbf{v}_-^k \\
	\mathbf{v}_+^k\cdot \mathbf{v}_-^k & \| \mathbf{v}_-^k \|^2
	\end{pmatrix}^{-1}
	\begin{pmatrix} 1 \\ 0 \end{pmatrix}.
	\end{aligned}
	\end{equation}
	Note that $\mathbf{n}_{LS}^N = \mathbf{n}_{LS}$.
	
	In the following, we show that any \emph{from-k} readout vector $\mathbf{n}_{LS}^k$ with $k<N$ is not fully contained in the plane spanned by vectors $\mathbf{v}_\pm$.
	The \emph{from-k} readout vector $\mathbf{n}_{LS}^k$ is contained in the plane spanned by cropped vectors  $\mathbf{v}^k_\pm$.
	Consider now any vector $\mathbf{a}$ which is orthogonal to the reservoir trajectory plane spanned by $\mathbf{v}_\pm$.
	We have
	\begin{equation}
	0 = \mathbf{a} \cdot \mathbf{v}_\pm = \sum_{i=1}^N ( \mathbf{a} )_i ( \mathbf{v}_\pm )_i.
	\end{equation}
	We have that cropped vectors $\mathbf{v}_\pm ^k$ do overlap with
	vector $\mathbf{a}$, because:
	\begin{equation}
	\mathbf{a} \cdot \mathbf{v}_\pm ^k = \sum_{i=1}^k ( \mathbf{a} )_i ( \mathbf{v}_\pm )_i = -\sum_{i=k+1}^N ( \mathbf{a} )_i ( \mathbf{v}_\pm )_i \neq 0.
	\end{equation}
	As a result, the readout vector $\mathbf{n}^k_{LS}$ also has a nonzero overlap with $\mathbf{a}$.
	
	\subsection{Analysis of least-squares regression: norm}
	\label{app:LSnorm}
	
\begin{figure}[t]
		\centering
		\includegraphics{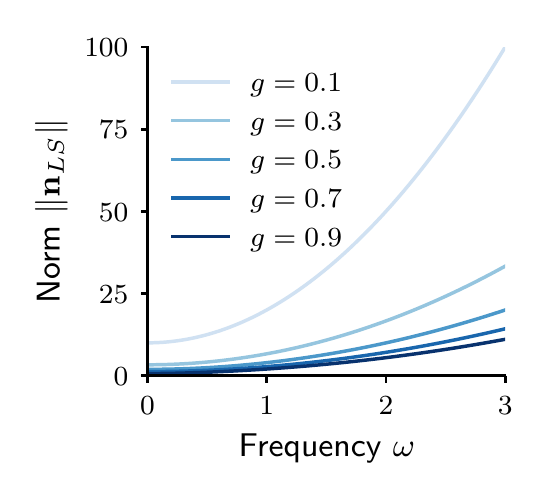}
		\caption{
			\textbf{Norm of LS-trained readout vector $\mathbf{n}$.}
			Computed via Eq.~\eqref{eq:nnorm} for a range of values $g$ and $\omega$.
			\label{fig:Norm_n}
		}
	\end{figure}

	We analytically compute the norm of the least-squares readout solution $\mathbf{n}_{LS}$ ($k=N$, Eqs.~\eqref{eq:readoutx} and \eqref{eq:readoutv}).
	We start from Eq.~\eqref{eq:readoutv} to write:
	\begin{equation}
	\begin{aligned}
	\mathbf{n}_{LS} &= 
	\begin{pmatrix}
	\mathbf{v}_+ & \mathbf{v}_-
	\end{pmatrix} 
	\left(C^R\right)^{-1}\begin{pmatrix} 1 \\ 0 \end{pmatrix} \\
	& = \frac{1}{\|\mathbf{v}_+\|^2\|\mathbf{v}_-\|^2(1-\cos^2(\theta))} \begin{pmatrix}
	\mathbf{v}_+ & \mathbf{v}_-
	\end{pmatrix} 
	\begin{pmatrix}
	\|\mathbf{v}_-\|^2 & - \mathbf{v}_+ \cdot \mathbf{v}_-\\
	- \mathbf{v}_+ \cdot \mathbf{v}_- & \|\mathbf{v}_+\|^2
	\end{pmatrix}
	\begin{pmatrix} 1 \\ 0 \end{pmatrix} \\
	&= \frac{1}{\|\mathbf{v}_+\|^2  \left(1-\text{cos}^2(\theta)\right)} \left( \mathbf{v}_+ - \frac{\mathbf{v}_+ \cdot \mathbf{v}_-}{\|\mathbf{v}_-\|^2}\mathbf{v}_-\right) \\
	&:= \frac{\|\mathbf{v}_+\|\|\mathbf{v}_-\|^2}{\text{det}\left(C^R\right)} \left( \hat{\mathbf{v}}_+ -\cos(\theta)\hat{\mathbf{v}}_-\right)
	\label{eq:readoutv2}
	\end{aligned}
	\end{equation}
	The vector within the parenthesis above is contained in the activity-spanning plane and is orthogonal to $\mathbf{v}_-$, and has norm
	\[
	\|\mathbf{v}_+\| \sqrt{1-\text{cos}^2(\theta)};
	\]
	the norm of $\mathbf{n}_{LS}$ is therefore given by
	\begin{equation}
	\|\mathbf{n}_{LS}\| = \frac{1}{N\|\mathbf{v}_+\| \sqrt{1-\text{cos}^2(\theta)}}  \label{eq:nnorm}
	\end{equation}
	We find that this expression is monotonically increasing in both $g$ and $\omega$, as shown in Supp.~Fig.~\ref{fig:Norm_n}.
	
	\subsection{Analysis of least-squares regression: outlier eigenvalues}
	\label{app:LS_eigs}
	
	In this section, we compute the outlier eigenvalues of $\bar{\mathbf{J}}$ which result from full LS regression ($k=N$, Eqs.~\eqref{eq:readoutx} and \eqref{eq:readoutv}).
	As derived in the main text, outliers obey:
	\begin{equation} \label{eq:no_oth_sol}
	\begin{split}
	1 &= {\mathbf{n}_{LS}}^\top [({\lambda} \mathbf{I} - \mathbf{J})^{-1} \mathbf{m}] := {\mathbf{n}_{LS}}^\top {\mathbf{x}}_\lambda.
	\end{split}
	\end{equation}
	Because of Eq.~\eqref{eq:decoding_fourier}, we know that the equation above admits the solutions $\lambda=\lambda_\pm = 1 \pm \omega$. In the following, we show that $\lambda_\pm$ is in fact the only solution admitted. To this end, we use Eq.~\eqref{eq:readoutv} to rewrite the equation as:
	\begin{equation}
	1 = \begin{pmatrix}
	1 & 0
	\end{pmatrix} 
	P^\top
	\begin{pmatrix}
	\mathbf{v}_+^\top \\ \mathbf{v}_-^\top
	\end{pmatrix} \mathbf{x}_{\lambda}
	\end{equation}
	where we defined the short-hand notation $P:=\left(C^R\right)^{-1}$. A little algebra yields:
	\begin{equation}
	\begin{split}
	1 &= \frac{1}{2}
	\begin{pmatrix}
	P_{11} & P_{21}
	\end{pmatrix}
	\begin{pmatrix}
	(\mathbf{x}_++\mathbf{x}_-)^\top \\ i(\mathbf{x}_+-\mathbf{x}_-)^\top
	\end{pmatrix}
	\mathbf{x}_{\lambda} \\
	&=\begin{pmatrix}
	P_{11} (\mathbf{x}_++\mathbf{x}_-)^\top + i P_{21} (\mathbf{x}_+-\mathbf{x}_-)^\top
	\end{pmatrix}
	\mathbf{x}_{\lambda} \\
	& = \frac{1}{2} \left[  \mathbf{x}_+^\top{\mathbf{x}}_\lambda  ( P_{11}+iP_{21} ) + \mathbf{x}_-^\top{\mathbf{x}}_\lambda  ( P_{11}-iP_{21} )  \right].
	\end{split}
	\end{equation}
	$P_{11}$ and $P_{21}$ are the elements of the first column of $P=\left(C^R\right)^{-1}$, namely:
	\begin{equation}
	\begin{split}
	&P_{11} = \frac{\|\mathbf{v}_- \|^2  }{\|\mathbf{v}_+\|^2\|\mathbf{v}_-\|^2(1-\cos^2(\theta))} \\
	&P_{21} = - \frac{\mathbf{v}_+\cdot\mathbf{v}_-}{\|\mathbf{v}_+\|^2\|\mathbf{v}_-\|^2(1-\cos^2(\theta))} .
	\end{split}
	\end{equation}
	
	We can evaluate dot products in the form:
	\begin{equation}
	\mathbf{x}_\pm^\top{\mathbf{x}}_\lambda = \left[({\lambda}_\pm \mathbf{I} - \mathbf{J})^{-1} \mathbf{m}  \right]^\top \left[({\lambda} \mathbf{I} - \mathbf{J})^{-1} \mathbf{m}  \right]
	\end{equation}
	by following Eq.~\eqref{eq:dot_prod_mean}, which was derived in Appendix \ref{app:spanning_vecs} by averaging over the random connectivity $\mathbf{J}$.
	This yields:
	\begin{equation} \label{eq:ridge_MDL}
	\begin{split}
	& 1 = \frac{N}{2} \left[  \frac{P_{11}+iP_{21}}{\lambda_+{\lambda}-g^2} + \frac{P_{11}-iP_{21}}{\lambda_-{\lambda}-g^2}  \right] 
	\end{split}
	\end{equation}
	which can be re-cast as a quadratic equation in $\lambda$:
	\begin{equation}
	(1+\omega^2){\lambda}^2 - \left[ 2g^2 + N (P_{11} + \omega P_{21}) \right] {\lambda}  +  g^4 + Ng^2P_{11} = 0.
	\end{equation}
	As we know that the quadratic equation above is satisfied by $\lambda = \lambda_\pm$, we conclude that Eq.~\eqref{eq:no_oth_sol} cannot admit other solutions beyond these two.

	\subsection{Condition number of the cross-correlation matrix}
	\label{app:PR_CN_extrema}
	
	The condition number of the reduced cross-correlation matrix $C^R$ is defined as:
	\begin{equation}
	c = \frac{\nu_1}{\nu_2}
	\end{equation}
	where $\nu_1$ and $\nu_2$ are the two eigenvalues of $C^R$. Their value can be computed as a function of the statistics of the spanning vectors $\mathbf{v}_\pm$, which in turns depend on $\omega$ and $g$ (see Appendix \ref{app:spanning_vecs}).
	
	In this section, we show that for fixed $g$, the condition number $c$ is minimized at the same value of $\omega$ which maximizes the participation ratio $d$; this frequency coincides with $\omega^*$ (see Appendix \ref{app:geometric_analysis}). Using:
	\begin{equation}
	\nu_{1,2} = \frac{1}{2}\left(  \gamma \pm \sqrt{\Delta^2} \right).
	\end{equation}
	we have:
	\begin{equation}
	d = 2\frac{\gamma^2}{\gamma^2 + \Delta^2} = \frac{2}{1 + (\frac{\Delta}{\gamma})^2},
	\end{equation}
	while
	\begin{equation}
	c = \frac{\gamma + \Delta}{\gamma - \Delta} = 1 + \frac{2}{\frac{\gamma}{\Delta} - 1}.
	\end{equation}
	The participation ratio $d$ is maximized when the quantity $\frac{\Delta}{\gamma}$ is minimized, which is precisely where the condition number $c$ is minimized.
	
	\subsection{Analysis of ridge regression}
	\label{app:ridge_regression}
	
	We start by computing the readout vector which performs ridge regression in the Fourier space.
	The ridge regressor of Eq.~\eqref{eq:decoding_fourier} can be written in terms of the $\mathbf{v}_\pm$ spanning vectors as \cite{HoerlRidge}:
	\begin{equation}
	\tilde{\mathbf{n}}_{R} = 
	\begin{pmatrix}
	\mathbf{v}_+ & \mathbf{v}_-
	\end{pmatrix}
	\tilde{P}
	\begin{pmatrix} 1 \\ 0 \end{pmatrix}
	\end{equation}
	where $\tilde{P} = \left(C^R + N \sigma^2 \mathbf{I}\right)^{-1}$ i.e.
	\begin{equation}
	\tilde{P} = \frac{1}{\det\left(C^R + N \sigma^2 \mathbf{I}\right)} \begin{pmatrix}
	\|\mathbf{v}_- \|^2 +N \sigma^2     &  -\mathbf{v}_+ \cdot \mathbf{v}_- \\
	-\mathbf{v}_+ \cdot \mathbf{v}_- & \|\mathbf{v}_+ \|^2 + N \sigma^2
	\end{pmatrix}.
	\end{equation}
	This yields the ridge regressor readout:
	\begin{equation}
	\tilde{\mathbf{n}}_R = \frac{\|\mathbf{v}_- \|^2\|\mathbf{v}_+ \|}{\det\left(C^R + N \sigma^2 \mathbf{I}\right)} \left[ \left(1 + \frac{N\sigma^2}{\|\mathbf{v}_- \|^2}\right) \mathbf{\hat{v}_+}  - \cos(\theta) \mathbf{\hat{v}_-} \right],
	\end{equation}
	where $\mathbf{\hat{v}}$ indicates a normalized vector.
	By comparison with the LS regressor, Eq.~\eqref{eq:readoutv2}, we observe that $\sigma$ has two effects on the readout: first, it reduces the norm, as expected from a regularizer.
	Second, it biases the readout vector towards $\mathbf{v}_+$.

	The outlier eigenvalues $\tilde{\lambda}_\pm$ imposed by the ridge regressor can be found by utilizing the same strategy as in Appendix \ref{app:LS_eigs}.
	We insert $P_{11}=\tilde{P}_{11}$,  $P_{21}=\tilde{P}_{21}$ and $\lambda_\pm = 1\pm\omega$ into Eq.~\eqref{eq:ridge_MDL} to obtain the equation for the outlier eigenvalues $\tilde{\lambda}$:
	\begin{equation} \label{eq:lambda_tilde_eq}
	(1+\omega^2)\tilde{\lambda}^2 - \left( 2g^2 + N\frac{-\omega\mathbf{v}_+\cdot\mathbf{v}_- + \|\mathbf{v}_- \|^2 + \sigma^2}{\det\left(C^R + N \sigma^2 \mathbf{I}\right)} \right) \tilde{\lambda}  +  \left( g^4 + g^2N\frac{\|\mathbf{v}_- \|^2 + \sigma^2}{\det\left(C^R + N \sigma^2 \mathbf{I}\right)} \right) = 0.
	\end{equation}
	
\begin{figure}	
		\centering
		
		\includegraphics{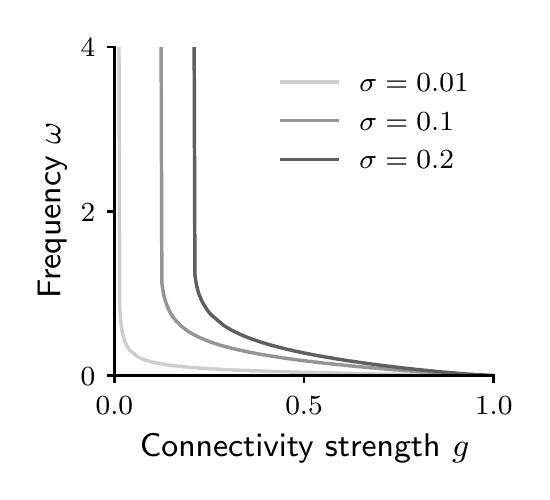}
		\caption{
			\textbf{Bifurcation diagram for networks trained through ridge regression.}
			For a range of the regularization parameter $\sigma$ (grey shades), the curve represents the location on the $g-\omega$ plane at which the outlier eigenvalues $\tilde{\lambda}_\pm$ bifurcate from two real eigenvalues into a pair or complex conjugate eigenvalues.
			\label{fig:bifurcation_diagram}
		}
	\end{figure}

	Depending on the values of $\sigma$, $\omega$ and $g$, the equation above admits real or complex conjugate eigenvalues (see Supp.~Fig.~\ref{fig:bifurcation_diagram}).
	For low frequencies, the dynamics are characterised by two real eigenvalues $\tilde{\lambda}_\pm$.
	As $\omega$ increases, a complex conjugate pair of eigenvalues is formed. 
	Importantly, their real part is always smaller than 1, yielding stable closed-loop dynamics.
	To see this, we approximate the real part of the solution to Eq.~\eqref{eq:lambda_tilde_eq} by assuming that $\sigma \ll 1$:
	\begin{equation}
	\mathcal{R}(\tilde{\lambda}_+) \approx \frac{g^2}{1+\omega^2} + \frac{\omega\mathbf{v}_+\cdot\mathbf{v}_- + \|\mathbf{v}_- \|^2 + \sigma^2}{2(1+\omega^2) \left[ \|\mathbf{v}_+ \|^2\|\mathbf{v}_-\|^2 \sin^2(\theta) + \sigma^2 \left( \|\mathbf{v}_+ \|^2 + \|\mathbf{v}_-\|^2 \right) \right] }.
	\end{equation}
	Now, by use of Eq.~\eqref{eq:geometric_qs} we evaluate
	\begin{equation*}
	\|\mathbf{v}_+ \|^2 + \|\mathbf{v}_-\|^2 = \frac{1}{1+\omega^2 - g^2},
	\end{equation*}
	implying that the pre-factor for the $\sigma^2$ term in the denominator is always larger than 1; as a consequence, $\sigma > 0$ always reduces the real part.
	
	To conclude, note that the analysis above allows to predict the behaviour of outlier eigenvalues when ridge regression is performed in the Fourier space (i.e.~from the 2-dimensional system of equations in Eq.~\eqref{eq:decoding_fourier}). In Figs.~\ref{fig:fig1} and \ref{fig:predicting_perf}, however, regression is performed in the temporal domain, on a higher-dimensional ($L'$-dimensional) set of equations (see Appendix \ref{app:training}). In order to compare the analytical prediction with trained networks, in Fig.~\ref{fig:predicting_perf} we thus scale the regularization parameter w.r.t.~the value of $\sigma$ which is used to derive analytical predictions, i.e.~we set $\left(\sigma^{\text{R}}\right)^2 = \sigma^2 \cdot L' / 2$ (see Appendix \ref{app:training}).

	\subsection{Characterization of the edge-of-chaos in non-linear networks}
	\label{app:EOC}
	
\begin{figure}
		
		\includegraphics[width=\textwidth]{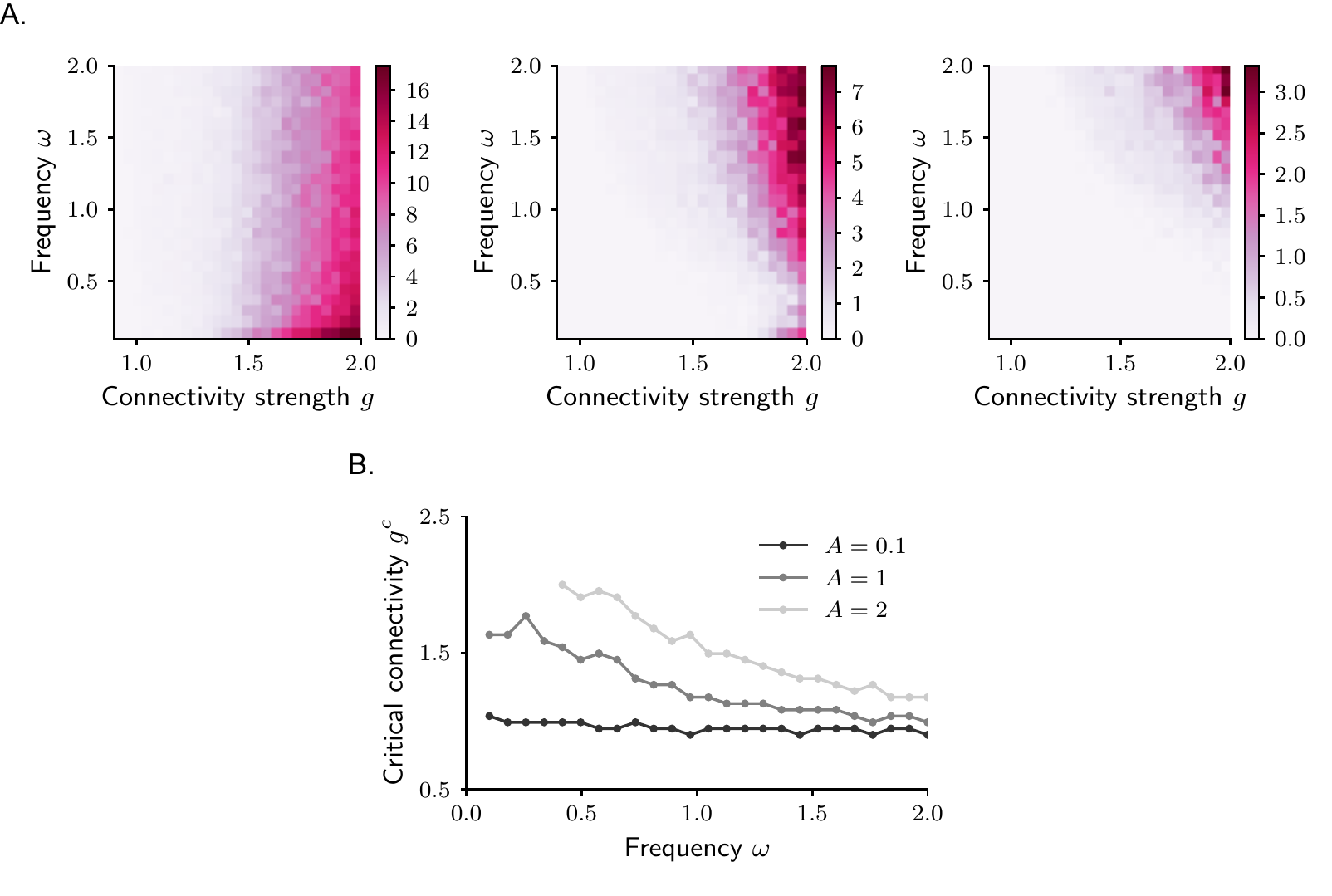}
		\caption{
			\textbf{Lyapunov dimensionality of driven non-linear networks.}
			{\bf A.} Lyapunov dimensionality (color code) of driven activity in non-linear networks as a function of connectivity strength $g$ and target frequency $\omega$.
			The target amplitude is taken to be $A=0.1$ (left), $A=1$ (center) and $A=2$ (right).
			{\bf B.} Edge-of-chaos curves extracted from the heat maps in {\bf A}: we plot the minimal value of connectivity strength $g^c$ for which the Lyapunov dimensionality is significantly non-zero. 
			\label{fig:EOC}
		}
	\end{figure}

	In analysing non-linear networks (Figs.~\ref{fig:fig1} and \ref{fig:fig6}), we varied the strength of internal connectivity $g$ in such a way that the open-loop dynamics driven by the target function $f(t)$ remains non-chaotic \cite{Sompolinsky1988} for every value of the forcing frequency tested. 
	The critical value of connectivity strength $g^c$ at which open-loop dynamics becomes chaotic depends on the target frequency $\omega$ and amplitude $A$ \cite{Rajan2010}, and was investigated numerically (Fig.~\ref{fig:EOC}).
	
	In order to find the critical values $g^c$, we start by computing the Lyapunov dimension $d_L$ of driven activity \cite{eckmann1985ergodic,geist1990comparison}, which is defined based on the Lyapunov spectrum $\Lambda = \lbrace \mu_i \rbrace_{i=1}^N$.
	Intuitively, Lyapunov exponents $\mu_i$ quantify the rate of exponential convergence or divergence of nearby trajectories along the different directions in state-space.
	If we order the exponents such that $\mu_1 \ge \mu_2 \ge \ldots \ge \mu_N$, the Kaplan-Yorke conjecture asserts that the Lyapunov dimension of the attractor is given by the index $j$ for which the number of contracting and expanding dimensions of the dynamics are balanced, namely the index $j$ for which 
	\begin{equation*}
	\sum_{i=1}^j \mu_i \ge 0,
	\end{equation*}
	and
	\begin{equation*}
	\sum_{i=1}^{j+1} \mu_i < 0.
	\end{equation*}
	In order to allow for attractors of fractal dimension, the dimension is defined as
	\begin{equation}
	d_L = j + \frac{\sum_{i=1}^j \mu_i}{| \mu_{j+1} |}.
	\end{equation}
	Note that $d_L$ is bound from below by $0$.  An attractor is considered chaotic iff the maximal Lyapunov exponent $\mu_1$ is positive, implying $d_L > 0$.
	
	The Lyapunov spectrum $\Lambda$ is computed numerically by evaluating the mean logarithmic growth of perturbations in the tangent space of the dynamics.
	To do so, we follow a QR-decomposition method, outlined in \cite{geist1990comparison}.
	The resulting Lyapunov dimension of activity in our driven reservoirs is show in Fig.~\ref{fig:EOC}{\bf A} for a range of $g$, $\omega$ and $A$ values. 
	From these maps we extract, for each value of the target amplitude $A$, the edge-of-chaos connectivity strength $g^c(\omega)$ which corresponds to the minimal value of $g$ for which $d_L > 0$ (Fig.~\ref{fig:EOC}{\bf B}).

\end{document}